\documentclass[10pt,twocolumn,letterpaper]{article}

\usepackage{cvpr}              % To produce the CAMERA-READY version
\usepackage[dvipsnames]{xcolor}

\definecolor{cvprblue}{rgb}{0.21,0.49,0.74}
\usepackage[pagebackref,breaklinks,colorlinks,citecolor=cvprblue]{hyperref}
\usepackage{indentfirst}
\usepackage{color}
\usepackage{multirow}
\usepackage{graphicx}
\usepackage{array}

\def\paperID{4257} % *** Enter the Paper ID here
\def\confName{CVPR}
\def\confYear{2024}

\title{Troublemaker Learning for Low-Light Image Enhancement}

\author{
Yinghao Song\textsuperscript{1,2}\quad
Zhiyuan Cao\textsuperscript{1,2}\quad
Wanhong Xiang\textsuperscript{3}\quad
Sifan Long\textsuperscript{1,2}\quad
Hongwei Ge\textsuperscript{4}\quad
Yanchun Liang\textsuperscript{1,2} \\
Bo Yang\textsuperscript{1,2}\quad
Chunguo Wu\textsuperscript{1,2}\\
\textsuperscript{1} {College of Computer Science and Technology, Jilin University, Jilin, China}\\
\textsuperscript{2} {Key Laboratory of Symbolic Computation and Knowledge Engineering of Ministry of Education},\\ 
{Jilin University, Jilin, China}\\
\textsuperscript{3} {YGSOFT INC}\\
\textsuperscript{4} {Dalian University of Technology}\\
{\tt\small \{songyh22, caozy20, longsf22\}@mails.jlu.edu.cn}, {\tt\small \{ycliang, yb, wcg\}@jlu.edu.cn}, \\
{\tt\small xiangwanhong@ygsoft.com},
{\tt\small hwge@dlut.edu.cn},
}

\begin{document}
\maketitle
% \footnotetext{\textsuperscript{\dag}Corresponding authors.} 
\begin{abstract}
\vspace{-0.3cm}
\label{sec:abstract}
Low-light image enhancement (LLIE) restores the color and brightness of underexposed images. Supervised methods suffer from high costs in collecting low/normal-light image pairs. Unsupervised methods invest substantial effort in crafting complex loss functions. We address these two challenges through the proposed \textbf{TroubleMaker Learning (TML)} strategy, which employs normal-light images as inputs for training. TML is simple: we first dim the input and then increase its brightness. TML is based on two core components. First, the troublemaker model (TM) constructs pseudo low-light images from normal images to relieve the cost of pairwise data. Second, the predicting model (PM) enhances the brightness of pseudo low-light images. Additionally, we incorporate an enhancing model (EM) to further improve the visual performance of PM outputs. Moreover, in LLIE tasks, characterizing global element correlations is important because more information on the same object can be captured. CNN cannot achieve this well, and self-attention has high time complexity. Accordingly, we propose \textbf{Global Dynamic Convolution (GDC)} with $O(n)$ time complexity, which essentially imitates the partial calculation process of self-attention to formulate elementwise correlations. Based on the GDC module, we build the UGDC model. Extensive quantitative and qualitative experiments demonstrate that UGDC trained with TML can achieve competitive performance against state-of-the-art approaches on public datasets. The code is available at \href{https://github.com/Rainbowman0/TML_LLIE}{https://github.com/Rainbowman0/TML\_LLIE}.
\end{abstract}

\section{Introduction}
\label{sec:intro}

\begin{figure}[t]
  \centering
  % \fbox{\rule{0pt}{2in} \rule{0.9\linewidth}{0pt}}
   \includegraphics[width=0.9\linewidth]{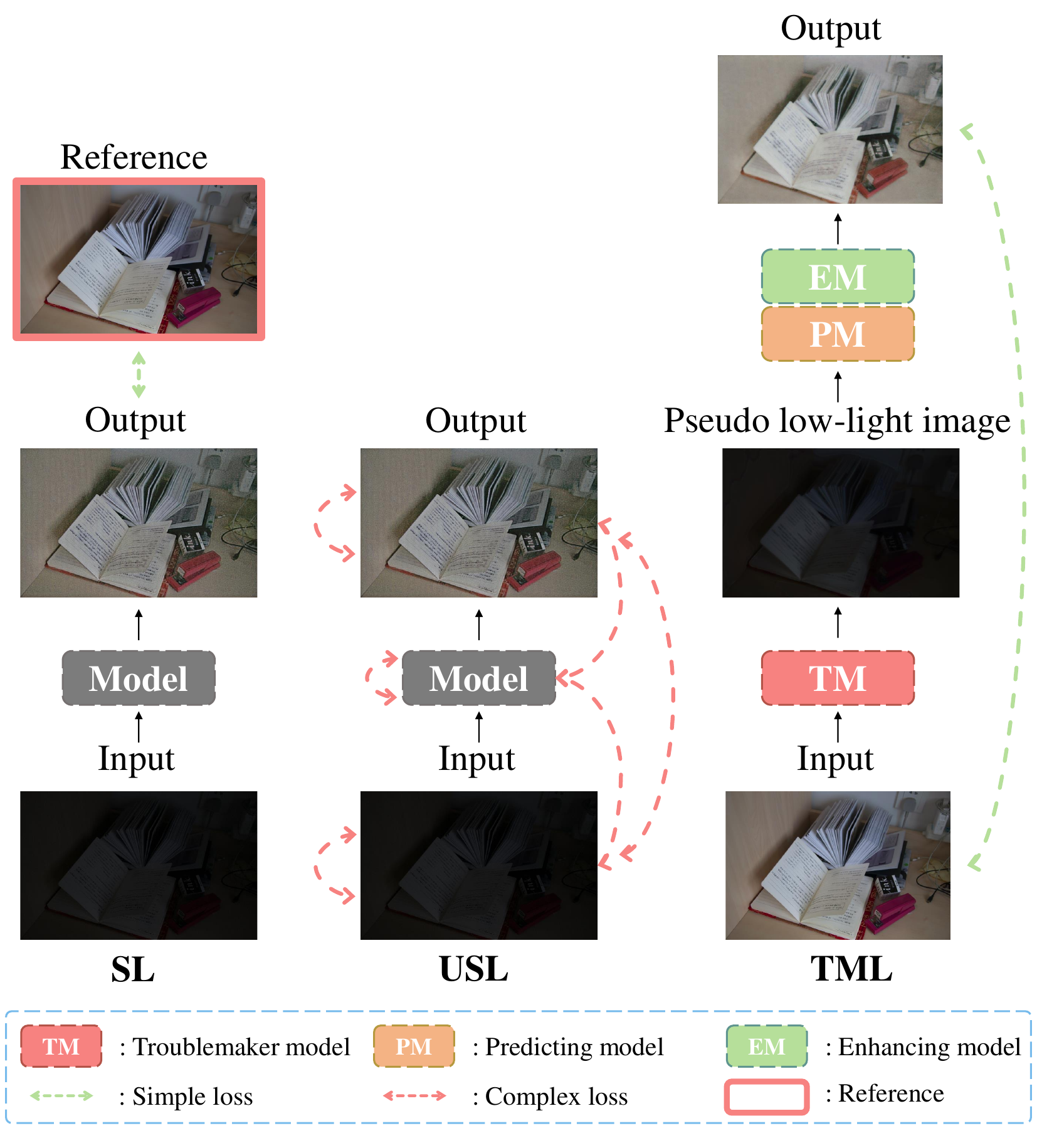}

   \caption{Three categories of training paradigms, SL, USL, and TML, denote supervised learning, unsupervised learning, and troublemaker learning, respectively. Supervised methods rely on paired data, while complex loss functions limit unsupervised methods. TML weakens the paired data restriction and exploits a simple loss function.}
   \label{fig01}
\end{figure}

Images taken in low-light environments suffer from underexposure and low contrast. They not only have poor visual effects but also create difficulties in downstream visual tasks~\cite{ref02,ref01}. Thus, low-light image enhancement (LLIE) aims at improving illumination and producing normal-light images. Traditional methods~\cite{ref03,ref08,ref04,ref05}, such as histogram equalization, adjust the contrast at different image locations. Retinex theory divides the image into two parts: illumination and reflection, which represent the lighting information of the environment and the reflection information of the object, respectively. More studies \cite{ref06,ref07,ref08,ref09} have utilized the Retinex theory~\cite{ref10} as it aligns with the human eye's color perception. However, traditional model-based methods require more handcrafted priors, thus restricting their flexibility. Problems such as color distortion and low clarity will occur in many dark-light scenes. With the increase of data volume and computing power, deep learning techniques have become prevalent.

From the perspective of the training paradigm, data-driven methods can be divided into two categories: supervised and unsupervised learning according to whether low/normal image pairs are necessary (SL, USL in \cref{fig01}). Supervised methods~\cite{ref11,ref14,ref15,ref16,ref19,ref20} combine outputs and references to calculate a simple loss function. However, it requires paired data and is expensive. Although unsupervised methods~\cite{ref21,ref22,ref25,ref37} remove the paired data restriction, they leverage inputs, outputs and intermediate features to jointly design complex loss functions and introduce many priors. Thus, we ask: \emph{how to design a training paradigm that both alleviates dependence on pairwise data and exploits a simple loss function?}

To answer this question, we propose the TroubleMaker Learning (TML in \cref{fig01}) strategy. TML is inspired by the adversarial concept of GAN~\cite{ref64} and introduces TM and PM to diminish and enhance image brightness, respectively. Specifically, TM learns the mapping from normal-light to low-light images. PM restores pseudo low-light images generated by TM into coarse normal-light images, and EM further optimizes the PM output in terms of color and brightness. Compared with other supervised and unsupervised methods, the most significant difference of TML is that it takes normal-light images as input, which reduces the reliance on paired data. TML can expand training data at a low cost and exploit a simple loss function. The testing process exclusively involves PM and EM.

In terms of model design, CNN is the most commonly used model structure in LLIE~\cite{ref14,ref28,ref30,ref31,ref16}. With the rise of the vision transformer~\cite{ref60}, there have been several works~\cite{ref17,ref18} using variations of the transformer. As the core transformer component, self-attention can characterize the global element correlations, which will enhance the quantitative and qualitative performance of LLIE methods. Current convolution-based structures cannot capture the correlations. Although vision transformer can achieve this, it has high computational complexity. Thus, we probe a problem: \emph{how to capture elementwise correlations with $O(n)$ time complexity?}

\begin{figure*}[t]
  \centering
  \includegraphics[width=1\linewidth]{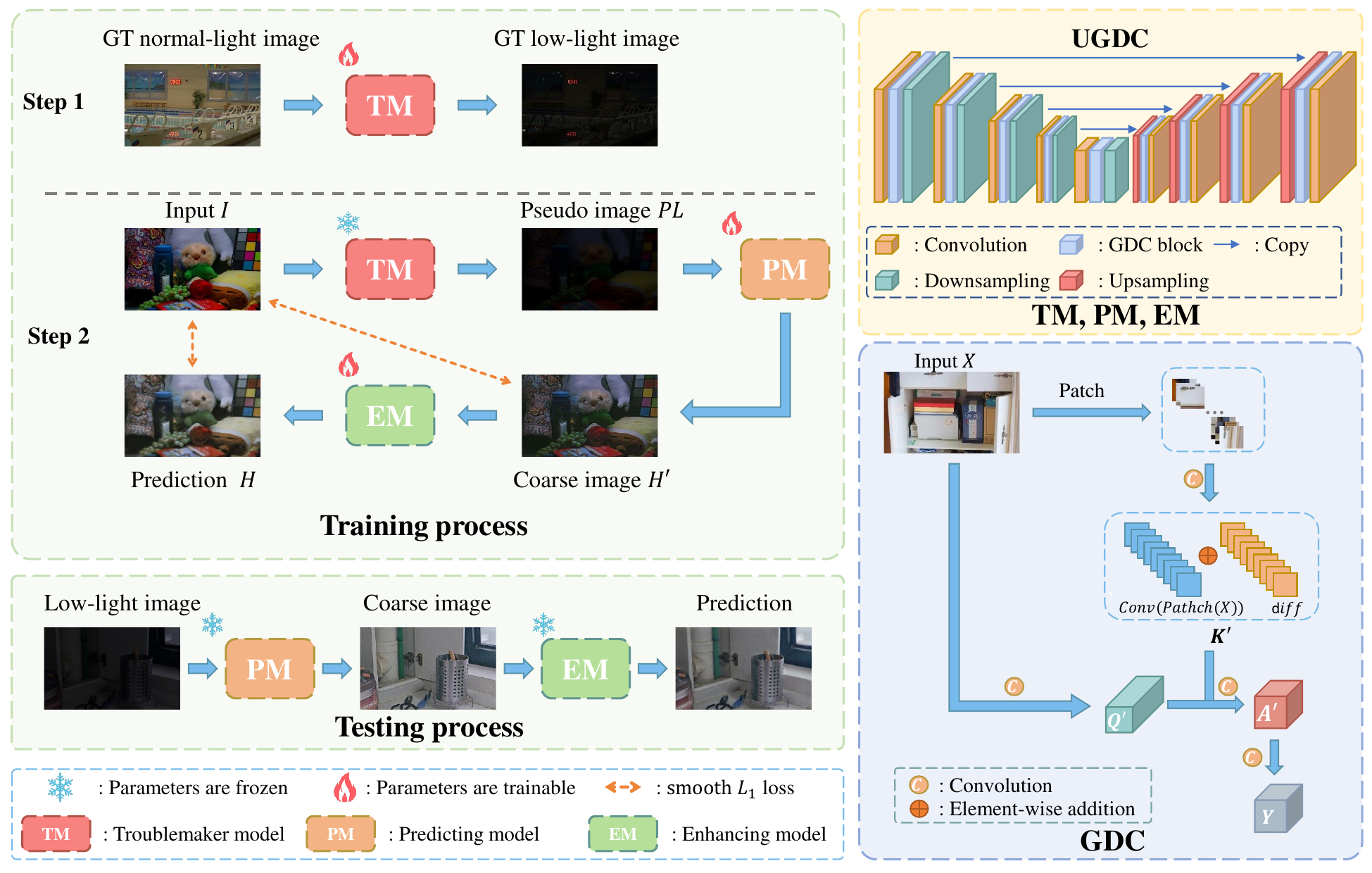}
   \caption{Overview of TML. The left side shows the TML training and testing process. In the training phase, TM only serves as a pseudo label generator. Step 1 requires a small quantity of paired data for training TM. Step 2 freezes TM and trains only PM and EM. The testing phase involves PM and EM. The UGDC (TM, PM and EM structure) and GDC details are on the right. GDC imitates the partial calculation process of self-attention and corresponds with \cref{eq04_2,eq05_2,eq06_2,eq07_2}.}
   \label{fig02}
\end{figure*}

The proposed Global Dynamic Convolution (GDC in \cref{fig02}) provides the answer. It is inspired by self-attention. In self-attention, the matrix multiplication of $Q$ and $K$ results in attention map $A$, whose value represents the relationship between elements at different positions. However, this calculation process is equivalent to $1\times1$ convolution (\cref{fig04_2}). GDC patches the input features and applies them as kernel parameters, which is called dynamic convolution. The output feature obtained by dynamic convolution is similar to attention map $A$ in self-attention, and its value also represents the elementwise correlations. Compared with other convolution-based models, the most significant advantage of GDC is that it can capture the correlation between elements while maintaining $O(n)$ time complexity.

Under the guidance of TML, the UGDC model (\cref{fig02}) we designed based on GDC achieved competitive results against the state-of-the-art approaches on public datasets. In summary, the contributions of our work are as follows:

\begin{itemize}[leftmargin=2em]
\medskip
\item
We propose the TML strategy that alleviates the dependence on pairwise data and complex priors.
\medskip
\item
We propose a novel convolution module GDC, which can capture the correlation between long-distance elements with $O(n)$ time complexity.
\medskip
\item
We conduct extensive qualitative, quantitative, and visual experiments to demonstrate the effectiveness of TML and GDC.
\end{itemize}

%-------------------------------------------------------------------------

\section{Related work}
\label{sec:related_work}

\subsection{Learning-based methods}

{ \bf Supervised methods.} LLNet~\cite{ref27} used MLP to lighten grayscale images. MBLLEN~\cite{ref14} proposed using CNN and multibranch structure to process low-light images. Chen \etal~\cite{ref28} first applied the U-Net structure to the LLIE field and achieved exciting results. Jiang \etal~\cite{ref29} dedicated to solving motion blur in low-light videos. Xu \etal~\cite{ref30} proposed to learn low-frequency and high-frequency information and then added the two to obtain the prediction result. DLN~\cite{ref31} proposed learning “brightness residuals”. The normal-light image was equal to the sum of the brightness residuals and the low-light image. Lv \etal~\cite{ref16} introduced exposure and noise attention to improve the prediction performance. Xu \etal~\cite{ref17} exploited CNN and transformer to model short-range and long-range dependencies simultaneously. LLFormer~\cite{ref18} designed the height and width axis attention to replace the multihead self-attention. Xu \etal~\cite{ref15} used the edge structure information of the image to obtain a clearer normal-light image.

In addition, there are some works based on Retinex theory. Retinex-Net~\cite{ref11} exploited Retinex theory earlier and proposed learning reflection and illumination. LightenNet~\cite{ref32} and DeepUPE~\cite{ref33} predicted the illumination of low-light images and then leveraged reflection as normal-light images. KinD~\cite{ref34} used a set of different exposure images instead of low-light and normal-light images during training. Recent works such as URetinex-Net~\cite{ref20} designed the initialization module so that the illumination and reflection initialization values are input-adaptive.

However, supervised methods require higher costs to collect paired data, so it is difficult to expand the training data.

{ \bf Unsupervised methods.} Unsupervised methods break the constraints of paired data. Inspired by the image processing software Photoshop, ExCNet~\cite{ref21} proposed learning "S-curve". ZeroDCE~\cite{ref22} was similar to ExCNet and guided the model to learn a more refined "S-curve". EnlightenGAN~\cite{ref35} exploited U-Net as a generator and used a double discriminator to learn global and local information. RUAS~\cite{ref23} established models to characterize the intrinsic underexposed structure of low-light images. RRDNet~\cite{ref36} decomposed the image into three parts: illumination, reflection and noise to achieve a better denoising effect. Recent works such as SCI~\cite{ref24} established a cascaded illumination learning process with weight sharing. NeRCo~\cite{ref25} introduced the idea of multimodality into LLIE for the first time. PairLIE~\cite{ref37} was based on Retinex theory and leveraged pairs of low-light images to train the model. It achieved competitive results. However, these works rely heavily on redundant loss functions to ensure convergence. They introduce many priors, which limits their generalization ability.

\subsection{CNN backbones}

U-Net is the most commonly used backbone in LLIE. It was first applied in biomedical image segmentation~\cite{ref38}. There is a series of works~\cite{ref11,ref14,ref29,ref30,ref16,ref15,ref35,ref39,ref40,ref18} that exploited U-Net-like structures as part of the model structure.

U-Net performs well in many fields because it integrates low/mid/high-level features. In fact, there are many excellent CNN structures in addition to U-Net. ResNet~\cite{ref41} first proposed using CNN to learn residuals. This idea was widely used in various fields. Inception~\cite{ref42} decomposed the convolution kernels so that the model had receptive fields of different sizes. DenseNet~\cite{ref43} is an extension of ResNet that proposed denser residual connections. MobileNet~\cite{ref44} leveraged depthwise and pointwise convolutions, greatly reducing the number of parameters and inference time. SENet~\cite{ref45} introduced the global attention mechanism into CNN and achieved competitive results. ConvNeXt~\cite{ref46} optimized the original ResNet by imitating the model structure and training strategy of Swin Transformer~\cite{ref61}. It proved that the potential of CNN has not been fully explored. Recent works such as DeepMAD~\cite{ref49} proposed a pure mathematical framework to maximize network entropy. FasterNet~\cite{ref50} designed partial convolution to reduce redundant information between channels. Xu \etal~\cite{ref17} and Wang \etal~\cite{ref18} illustrated that characterizing long-distance elementwise correlations are helpful in LLIE. Some efforts use convolution kernels for global attention to model long-range dependencies. RepLKNet~\cite{ref47} enlarges the kernel size to 31x31. VAN~\cite{ref48} employs depth-wise convolution, depth-wise dilation convolution, and pointwise convolution to expand the receptive field while reducing computational complexity. In contrast, GDC uses image patches as convolution kernel parameters to mimic Q-K multiplication in self-attention, thereby integrating global attention.
\section{Method}
\label{sec:method}

In this section, we first introduce TML, including motivation, overall process, and detailed analysis. The GDC is then presented, including motivation and design details.

\subsection{Troublemaker learning}
\label{sec:method01}

{\bf Motivation.} TML is inspired by the adversarial concept from GAN~\cite{ref64} and introduces the "bad student" (\ie TM) and "good student" (\ie PM). Specifically, TM intentionally learns a poor feature, which is a pseudo low-light image based on a normal-light input. The pseudo low-light image serves as a pseudo label for subsequent training. In contrast, PM learns the mapping from low-light to normal-light images, enhancing image brightness.

{\bf TML overall process.} The TML pipeline is illustrated in \cref{fig02}. The training process consists of two steps. Step 1 is to obtain a pseudo low-light image generator; thus, a small number of low/normal image pairs are required to train TM. It is worth noting that we use normal-light images as inputs and low-light images as labels. The input of step 2 is a normal-light image $I$. TM first reduces its brightness to generate a pseudo low-light image $PL$. Then, PM increases the $PL$ brightness to obtain a coarse normal-light image $H'$. Finally, EM further optimizes the brightness and color of $H'$ to obtain normal-light prediction $H$. In step 2, we freeze TM, update the PM and EM parameters, and the test process includes only PM and EM. TM, PM, and EM all employ the UGDC network architecture (top right of \cref{fig02}).

We consider two EM design options. In one option, EM directly predicts the mapping from $H'$ to $H$:

\begin{equation}
  H = EM(H')
  \label{eq01}
\end{equation}

The other option predicts the residual between $H'$ and $H$, prediction $H$ is the difference between $H'$ and the residual:

\begin{equation}
  residual = EM(H')
  \label{eq02}
\end{equation}
\begin{equation}
  H = H' - residual
  \label{eq03}
\end{equation}

Ablation experiments confirmed that the latter performs better; details are shown in \cref{tab02v2}. In step 2, PM and EM are trained separately. The distance between $H'$ and $I$ and between $H$ and $I$ is measured by the $smooth_{\mathcal{L}_1}$~\cite{ref56} loss function:

\begin{equation}
  \mathcal{L} = \frac{1}{n} \displaystyle\sum_{i} smooth_{\mathcal{L}_1}({y}_i, {h}_i) 
  \label{eq04v1}
\end{equation}

\begin{equation}
  smooth_{\mathcal{L}_1}({y}_i, {h}_i) = \begin{cases} 
  0.5({y}_i - {h}_i)^2 &\text{if } |{y}_i - {h}_i|<1 \\
  |{y}_i - {h}_i| - 0.5 &\text{otherwise}
  \end{cases}
  \label{eq04v2}
\end{equation}

where $y_i$ is the element in the input $I$ in step 2, $h_i$ represents the element in $H$ and $H'$, and $n$ represents the number of pixels. For a $400\times640$ RGB image, $n = 400 \times 640 \times 3 = 768,000$.

\textbf{TML alleviates reliance on pairwise data.} Although step 1 still requires paired data, experiments have proven that a good pseudo low-light generator can be obtained with only 200 pairwise data (\cref{tab_TM_ablation}). The testing process includes only PM and EM. It is worth noting that only normal-light images are used when training both, which greatly weakens the paired data constraints. Therefore, the training data can be expanded at a low cost to improve the generalization of the model.

\begin{figure}[t]

  \begin{subfigure}{0.32\linewidth}
    \includegraphics[width=0.98\linewidth]{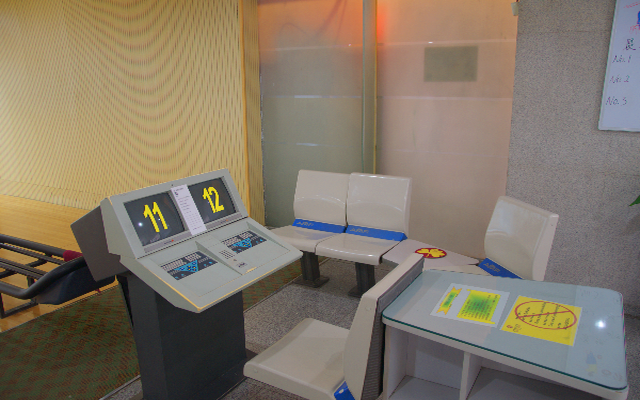}
  \end{subfigure}
  \hfill
  \begin{subfigure}{0.32\linewidth}
    \includegraphics[width=0.98\linewidth]{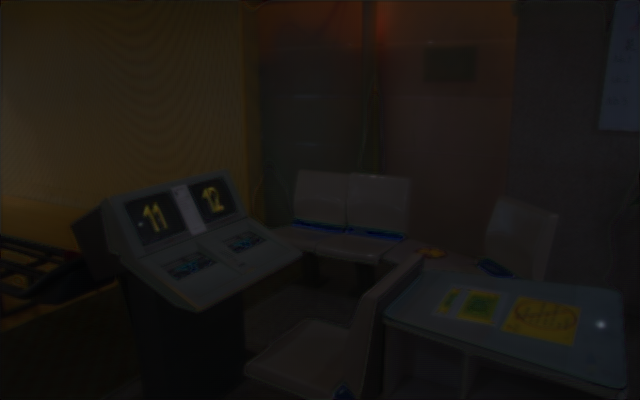}
  \end{subfigure}
  \hfill
  \begin{subfigure}{0.32\linewidth}
    \includegraphics[width=0.98\linewidth]{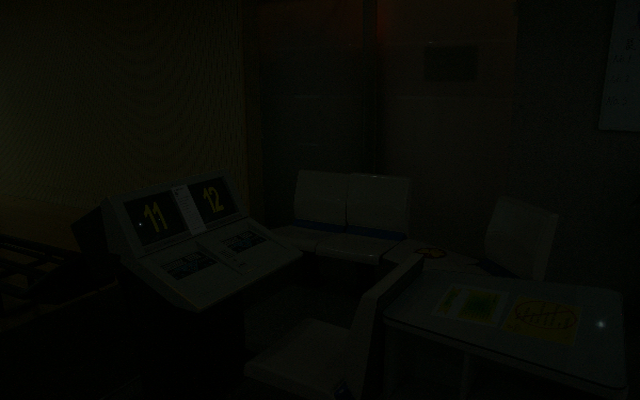}
  \end{subfigure}
  \vspace{0.5mm}

  \begin{subfigure}{0.32\linewidth}
    \includegraphics[width=0.98\linewidth]{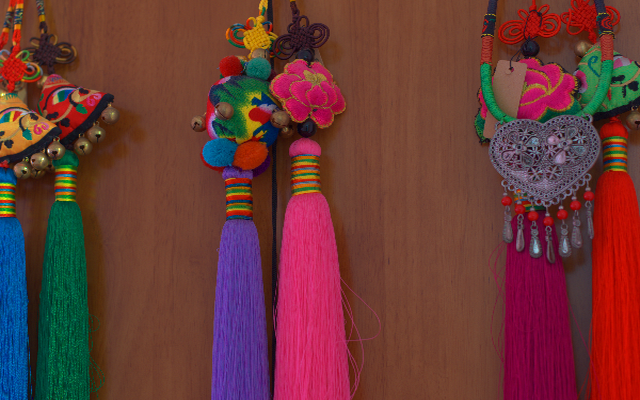}
  \end{subfigure}
  \hfill
  \begin{subfigure}{0.32\linewidth}
    \includegraphics[width=0.98\linewidth]{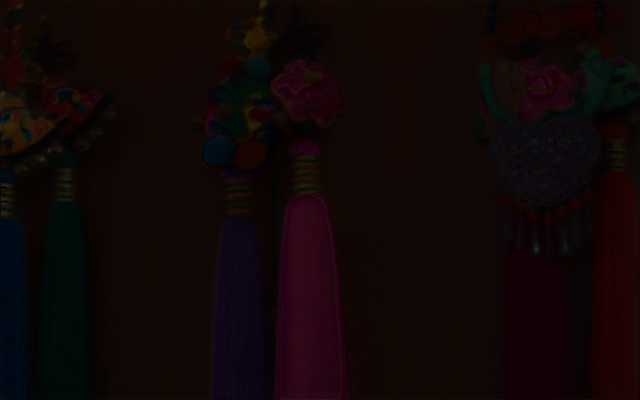}
  \end{subfigure}
  \hfill
  \begin{subfigure}{0.32\linewidth}
    \includegraphics[width=0.98\linewidth]{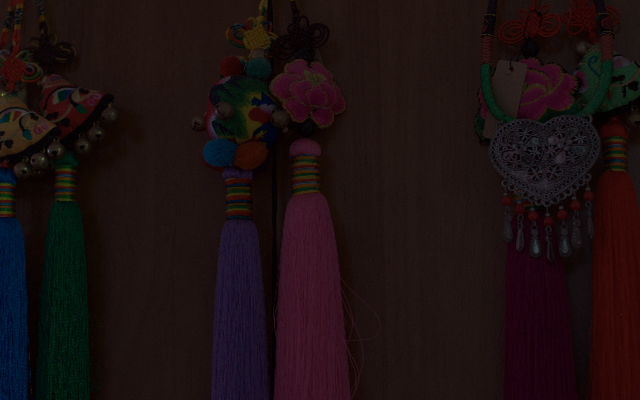}
  \end{subfigure}
  \vspace{0.5mm}

  \begin{subfigure}{0.32\linewidth}
    \includegraphics[width=0.98\linewidth]{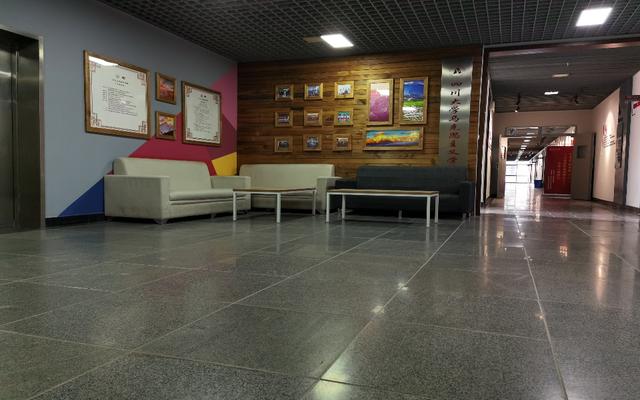}
  \end{subfigure}
  \hfill
  \begin{subfigure}{0.32\linewidth}
    \includegraphics[width=0.98\linewidth]{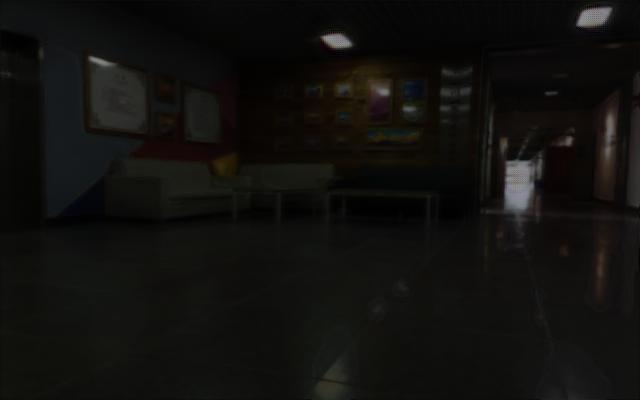}
  \end{subfigure}
  \hfill
  \begin{subfigure}{0.32\linewidth}
    \includegraphics[width=0.98\linewidth]{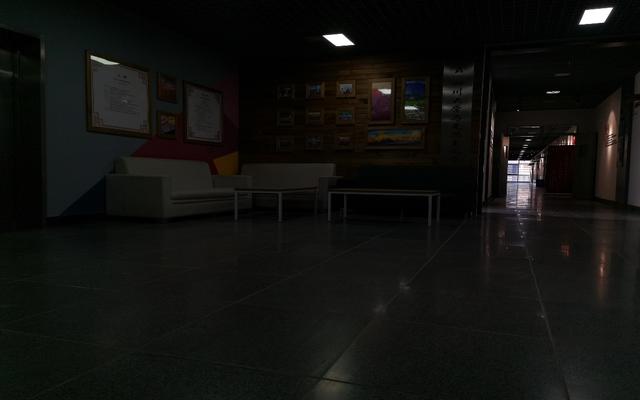}
  \end{subfigure}
  \vspace{0.5mm}

  \begin{subfigure}{0.32\linewidth}
    \includegraphics[width=0.98\linewidth]{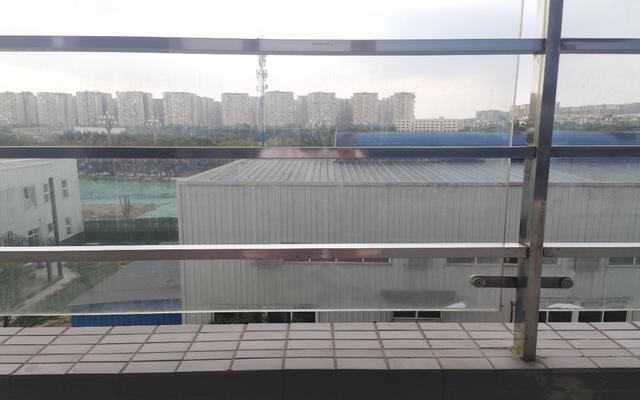}
    \caption*{Normal-light images}
  \end{subfigure}
  \hfill
  \begin{subfigure}{0.32\linewidth}
    \includegraphics[width=0.98\linewidth]{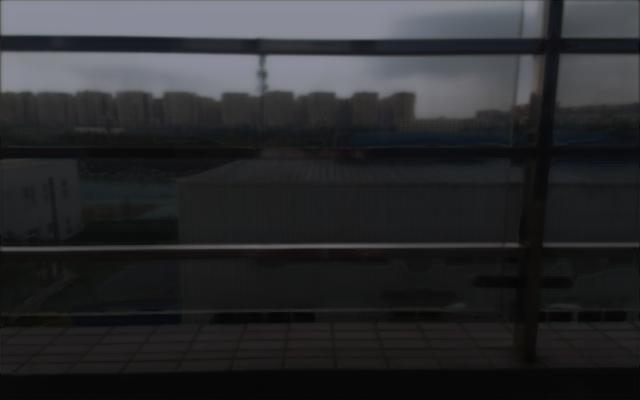}
    \caption*{TM predictions}
  \end{subfigure}
  \hfill
  \begin{subfigure}{0.32\linewidth}
    \includegraphics[width=0.98\linewidth]{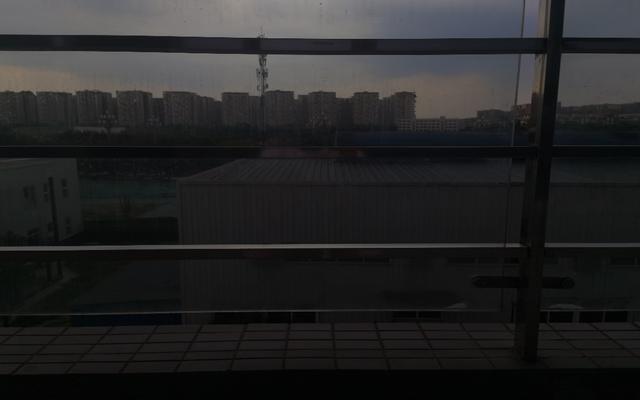}
    \caption*{Real low-light images}
  \end{subfigure}
  \vspace{0.5mm}

  \caption{Pseudo low-light images. The first and third columns are normal-light and real low-light images obtained via adjusting ISO. The second column is the pseudo low-light images predicted by TM.}
  \label{fig03_2}
\end{figure}

Like most deep learning methods, TML is data-driven. Previous studies have shown that data-driven techniques can create realistic normal-light images from low-light RGB images. Thus, we believe TM can similarly produce realistic low-light images from normal-light RGB images. \cref{tab_TM_ablation} demonstrates the similarity in data distribution between TM-generated and real low-light images. The prediction results of TM are illustrated in \cref{fig03_2}. Traditional supervised methods exploit low/normal-light image pairs. To ensure the same semantic information, a common approach is to obtain two images with different exposures via adjusting the camera ISO~\cite{ref11,ref12}. However, this sometimes does not guarantee the consistency of semantic information. As shown in the fourth row of \cref{fig03_2}, we find that the sky colors of low-light and normal-light images are different because low-light images have lower ISO and, therefore, retain more color details. TML can address this issue. The two have stronger semantic consistency because pseudo low-light images are directly predicted by normal-light images. In addition, we observe that TML predictions are more blurry than real low-light images. This is equivalent to a data augmentation strategy, so the trained model is more robust.

\subsection{Global dynamic convolution}
\label{sec:method02}

\begin{figure}[t]
  \centering
  % \fbox{\rule{0pt}{2in} \rule{0.95\linewidth}{0pt}}
   \includegraphics[width=0.98\linewidth]{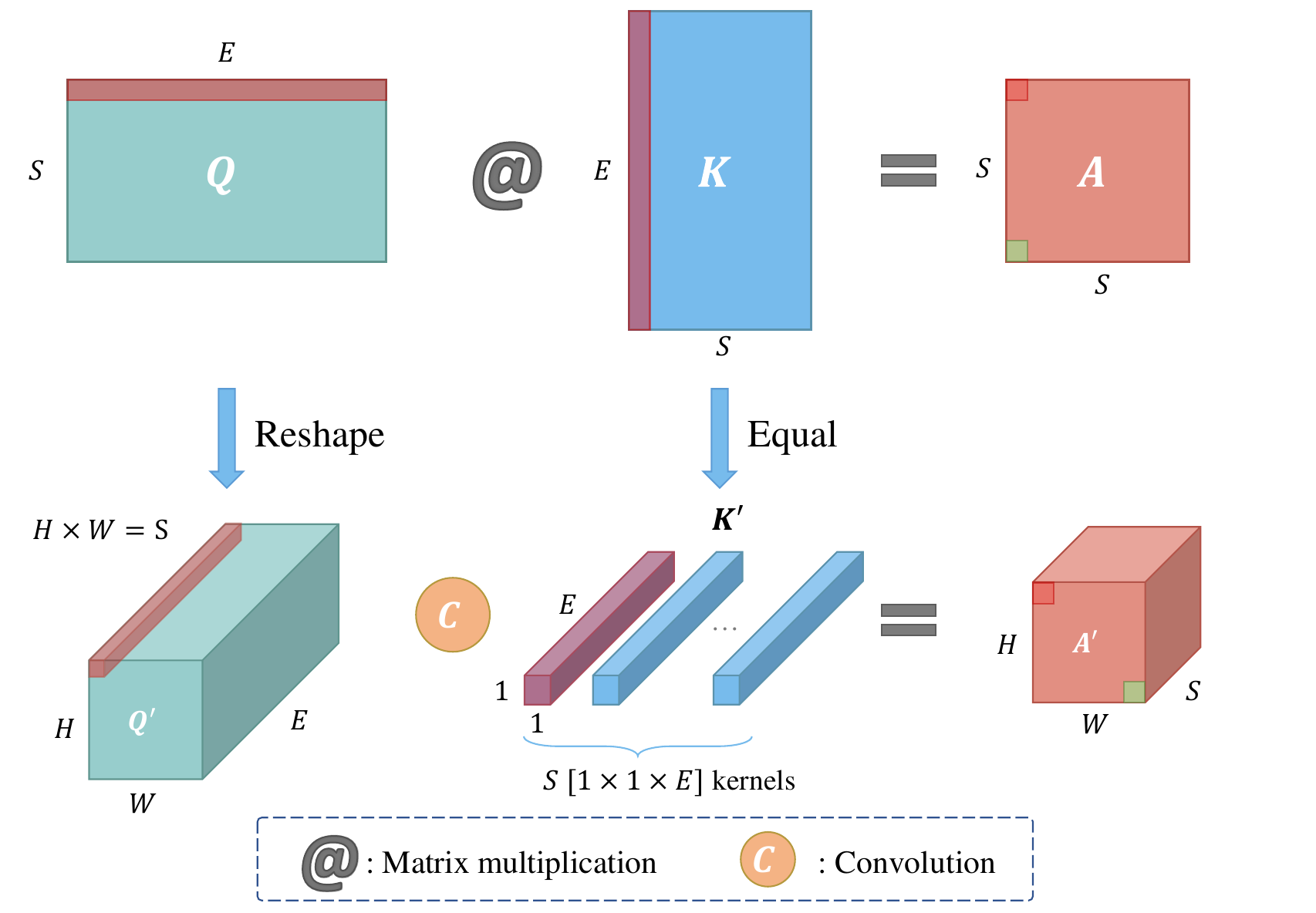}

   \caption{The matrix multiplication between $Q$ and $K$ in self-attention (top) can be replaced by a convolution operation (bottom). We use red and green to mark the equivalent elements in $Q$ and $Q'$, $K$ and $K'$, $A$ and $A'$.}
   \label{fig04_2}
\end{figure}

{\bf Motivation.} CNN has inductive biases such as locality and translation invariance and retains the 2D image structure information. However, it cannot capture elementwise correlations. Vision transformers can achieve this; however, the computational complexity is high. Recent works~\cite{ref17,ref18} have demonstrated that characterizing correlations between elements improves model performance for darker regions.

{\bf Global dynamic convolution.} The self-attention calculation process can be expressed by:
\begin{equation}
  A = {Q}{\times}K^T
  \label{eq02_2}
\end{equation}
\begin{equation}
  Y = \mathit{softmax}(\frac{A}{\sqrt{d_k}})\times{V}
  \label{eq03_2}
\end{equation}
where $Q$, $K$ and $V$ are attained by the linear transformation of the input. $A$ refers to attention map, which represents the correlations between elements at different positions. $\mathit{d}_{k}$ represents the dimensions of $K$.

GDC applies convolution to imitate \cref{eq02_2}. As illustrated in \cref{fig04_2}, we reshape $Q$ into a 3-dimensional feature map $Q'$, and regard $K$ as $K'$, which contains S $1\times{1}\times{E}$ kernels. Accordingly, matrix multiplication between $Q$ and $K$ is equivalent to convolution between $Q'$ and $K'$. Each element of $A'$ also represents the correlation between different elements of $Q'$ and $K'$.

It is worth noting that in self-attention, parameters $Q$ and $K$ are input-adaptive. However, if we regard $K'$ as S ${1}\times{1}\times{E}$ kernels, the parameters of$K'$ are fixed. To address this, we propose GDC block (the right part of \cref{fig02}).

The calculation process of GDC can be expressed as
\begin{equation}
  K'=Conv(Patch(X)) + \mathit{diff}
  \label{eq04_2}
\end{equation}
\begin{equation}
  Q'=Conv(X)
  \label{eq05_2}
\end{equation}
\begin{equation}
  A'=Conv_{K'}(Q')
  \label{eq06_2}
\end{equation}
\begin{equation}
  Y=Conv(A')
  \label{eq07_2}
\end{equation}

where $Conv_{K'}(Q')$ means applying $K'$ as the convolution kernels and convolving with $Q'$. $\mathit{Patch}$ refers to breaking the image into small patches, and $\mathit{diff}$ refers to the trainable parameters.

GDC first patches the input $X$ and then performs convolution. The obtained result is added to the trainable parameter $\mathit{diff}$ to increase the nonlinearity of $K'$. $K'$ and $Q'$ are convolved to obtain $A'$. The result $Y$ is obtained by convolution transformation of $A'$. We replaced several convolutional layers in U-Net~\cite{ref38} with GDC block and designed UGDC (the right part of \cref{fig02}).

{\bf Advantages of GDC.} GDC applies convolution to imitate the matrix multiplication process of $Q$ and $K$ in self-attention. Therefore, unlike previous convolutional modules, GDC can capture the element-wise correlations between long-distance (\cref{fig11}) and the parameters are input-adaptive. In addition, the time complexity of self-attention is $O(n^2)$, while GDC is $O(n)$, which means that the time complexity increases linearly as the image size increases, so it is more economical.

\section{Experiments}
\label{sec:exp}

\begin{figure*}[t]
  \centering
  \begin{subfigure}{0.12\linewidth}
    \includegraphics[width=1\linewidth]{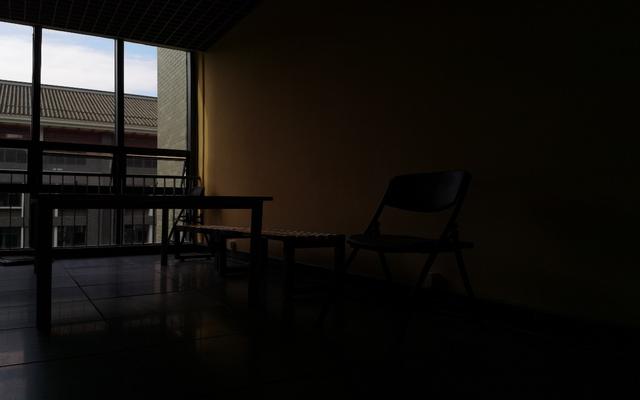}
    \caption*{Input}
  \end{subfigure}
  \hfill
  \begin{subfigure}{0.12\linewidth}
    \includegraphics[width=1\linewidth]{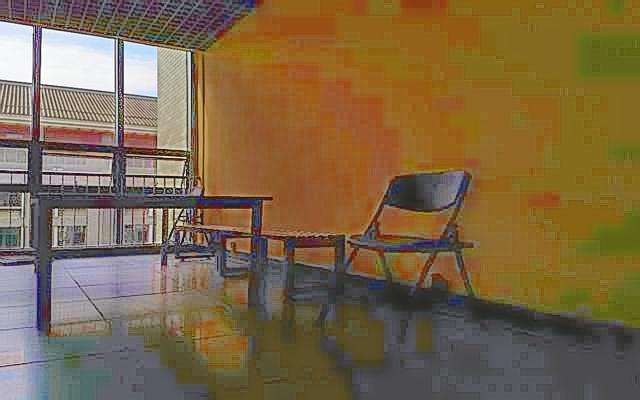}
    \caption*{Retinex}
  \end{subfigure}
  \hfill
  \begin{subfigure}{0.12\linewidth}
    \includegraphics[width=1\linewidth]{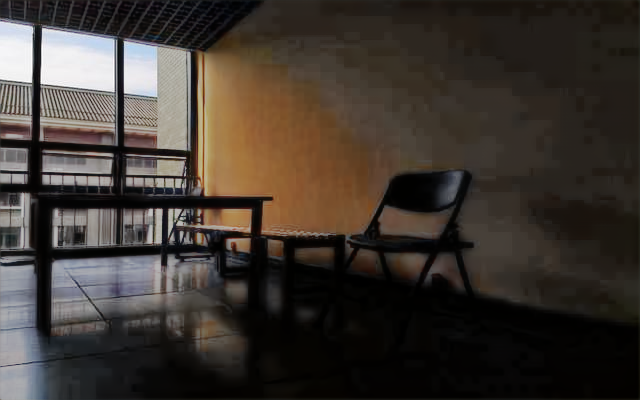}
    \caption*{KinD}
  \end{subfigure}
  \hfill
  \begin{subfigure}{0.12\linewidth}
    \includegraphics[width=1\linewidth]{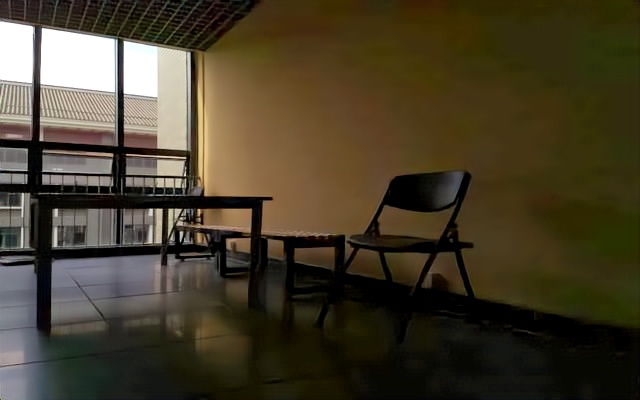}
    \caption*{MBLLEN}
  \end{subfigure}
  \hfill
  \begin{subfigure}{0.12\linewidth}
    \includegraphics[width=1\linewidth]{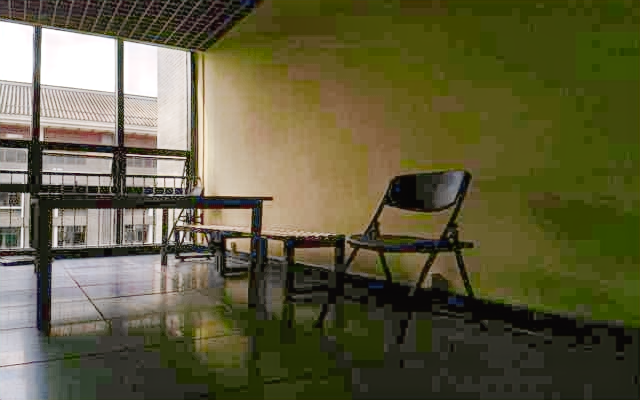}
    \caption*{GLAD}
  \end{subfigure}
  \hfill
  \begin{subfigure}{0.12\linewidth}
    \includegraphics[width=1\linewidth]{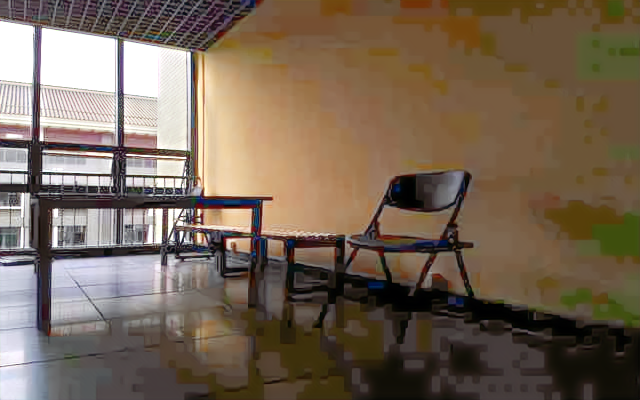}
    \caption*{URetinex}
  \end{subfigure}
  \hfill
  \begin{subfigure}{0.12\linewidth}
    \includegraphics[width=1\linewidth]{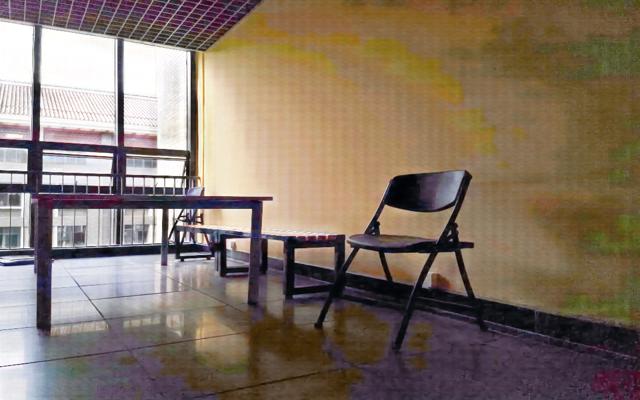}
    \caption*{SNR}
  \end{subfigure}
  \hfill
  \begin{subfigure}{0.12\linewidth}
    \includegraphics[width=0.98\linewidth]{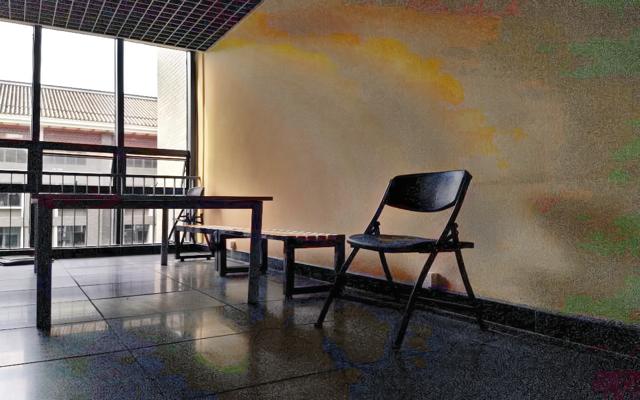}
    \caption*{LLFlow}
  \end{subfigure}

  \begin{subfigure}{0.12\linewidth}
    \includegraphics[width=1\linewidth]{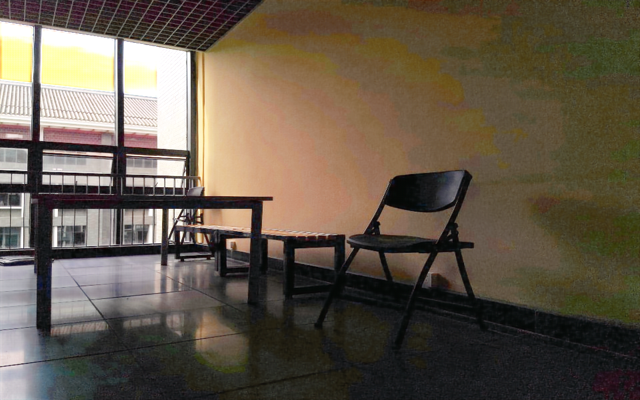}
    \caption*{Retinexformer}
  \end{subfigure}
  \hfill
  \begin{subfigure}{0.12\linewidth}
    \includegraphics[width=1\linewidth]{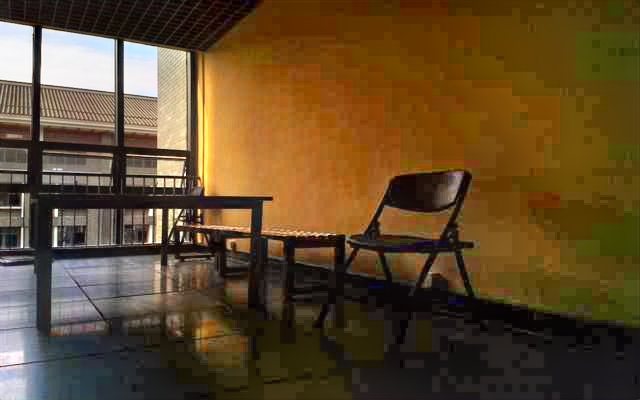}
    \caption*{EnGAN}
  \end{subfigure}
  \hfill
  \begin{subfigure}{0.12\linewidth}
    \includegraphics[width=1\linewidth]{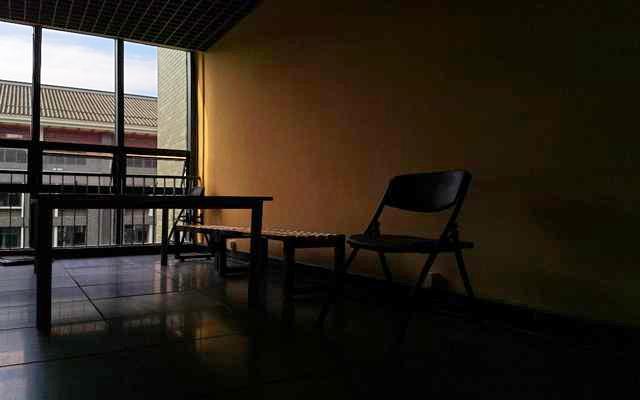}
    \caption*{RRDNet}
  \end{subfigure}
  \hfill
  \begin{subfigure}{0.12\linewidth}
    \includegraphics[width=1\linewidth]{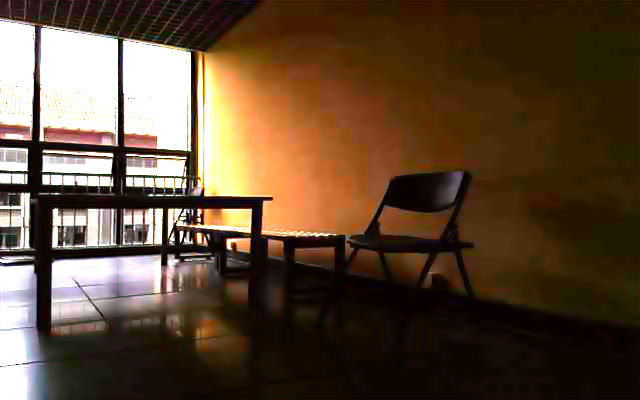}
    \caption*{RUAS}
  \end{subfigure}
  \hfill
  \begin{subfigure}{0.12\linewidth}
    \includegraphics[width=1\linewidth]{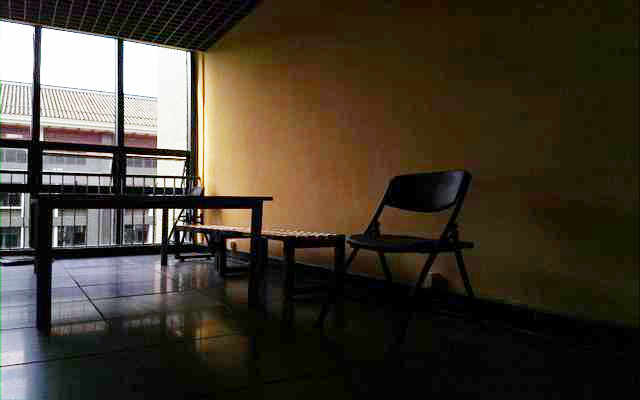}
    \caption*{SCI}
  \end{subfigure}
  \hfill
  \begin{subfigure}{0.12\linewidth}
    \includegraphics[width=1\linewidth]{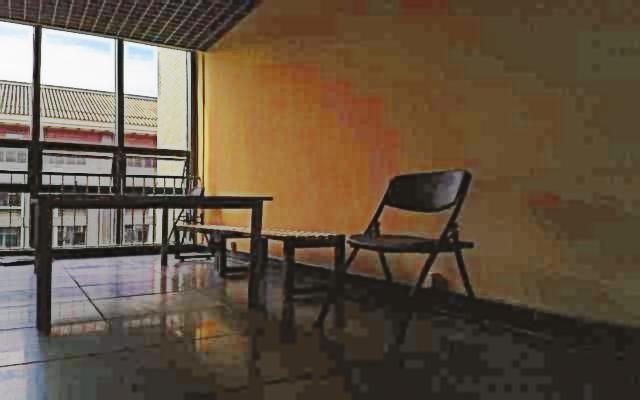}
    \caption*{PairLIE}
  \end{subfigure}
  \hfill
  \begin{subfigure}{0.12\linewidth}
    \includegraphics[width=1\linewidth]{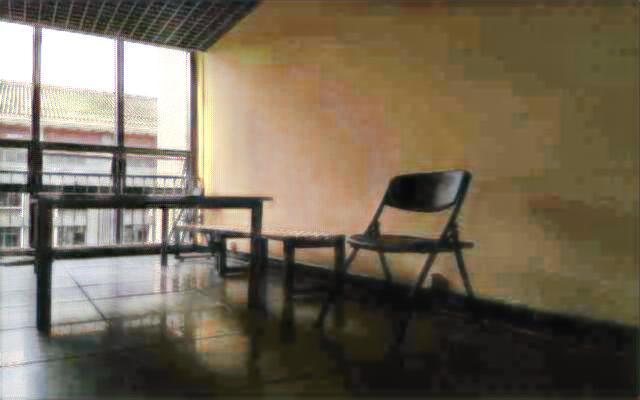}
    \caption*{TML (Ours)}
  \end{subfigure}
  \hfill
  \begin{subfigure}{0.12\linewidth}
    \includegraphics[width=1\linewidth]{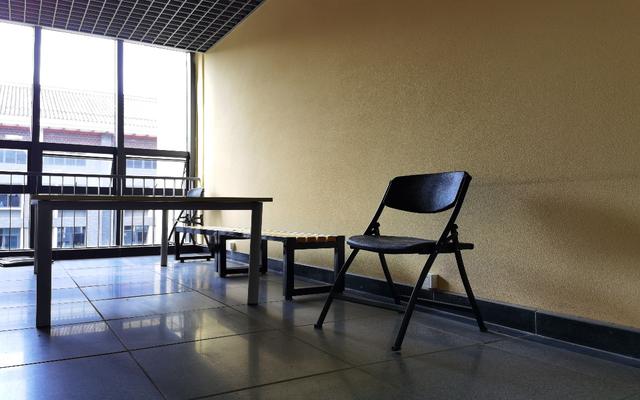}
    \caption*{Reference}
  \end{subfigure}

  \caption{Qualitative comparisons with state-of-the-art methods. TML makes 
reasonable trade-offs in brightness, color, and image quality.}
  \label{fig06}
\end{figure*}

\begin{table*}[t]
\renewcommand\arraystretch{1.1}
\centering
\resizebox{2\columnwidth}{!}{
\begin{tabular}{c|c|cccc|cccc|cccc}
\hline
\multirow{2}{*}{Method} & \multirow{2}{*}{Type} & \multicolumn{4}{c|}{LSRW}                                           & \multicolumn{4}{c|}{LOL}                                          & \multicolumn{4}{c}{RELLISUR}                                       \\ \cline{3-14} 
                        &                       & PSNR$\uparrow$            & SSIM$\uparrow$           & LPIPS$\downarrow$          & NIQE$\downarrow$           & PSNR$\uparrow$            & SSIM$\uparrow$           & LPIPS$\downarrow$          & NIQE$\downarrow$           & PSNR$\uparrow$            & SSIM$\uparrow$           & LPIPS$\downarrow$          & NIQE$\downarrow$           \\ \hline
Retinex~\cite{ref11}                 & S                     & 15.885          & 0.652          & 0.388          & 5.652          & 17.575          & 0.775          & 0.382          & 4.201          & 15.945          & 0.289          & 0.532          & 4.643          \\
KinD~\cite{ref19}                    & S                     & 16.989          & 0.674          & 0.330          & 4.659          & 17.755          & 0.841          & 0.176          & 4.411          & 16.307          & 0.304          & 0.425          & 3.948          \\
MBLLEN~\cite{ref14}                  & S                     & 16.930          & 0.663          & 0.342          & 5.341          & 17.857          & 0.790          & 0.226          & 4.395          & 16.673          & 0.429          & 0.400          & 4.187          \\
GLAD~\cite{ref59}                    & S                     & 18.763          & 0.725          & 0.306          & 4.524          & 19.743          & 0.805          & 0.327          & 5.510          & 16.636          & 0.349          & 0.472          & 4.050          \\
URetinex~\cite{ref20}                & S                     & 17.682          & 0.702          & 0.303          & 4.875          & 20.162          & \textbf{0.876} & 0.131          & 4.253          & 18.221          & 0.401          & 0.332          & 5.632          \\
SNR~\cite{ref65}                     & S                     & 17.733          & 0.626          & 0.234          & 4.347          & 24.639          & 0.861          & \textbf{0.114} & 4.840          & 17.791          & 0.384          & 0.387          & 5.559          \\
LLFlow~\cite{ref67}                  & S                     & 17.976          & 0.621          & \textbf{0.213} & \textbf{3.878}          & 20.941          & 0.863          & 0.117          & 5.662 & 18.429          & 0.389          & 0.337          & 4.136          \\
Retinexformer~\cite{ref66}           & S                     & 18.201          & 0.613          & 0.220          & 3.912 & \textbf{21.794} & 0.858          & 0.130          & \textbf{3.805}          & 18.810          & 0.342          & \textbf{0.317} & 4.876          \\ \hline
EnGAN~\cite{ref35}                   & U                     & 17.854          & 0.699          & 0.330          & 4.274          & 17.595          & 0.772          & 0.331          & 4.247          & 17.063          & 0.362          & 0.448          & 4.485          \\
RRDNet~\cite{ref36}                  & U                     & 13.494          & 0.542          & 0.338          & 6.082          & 11.350          & 0.545          & 0.370          & 6.241          & 10.867          & 0.163          & 0.505          & 4.841          \\
RUAS~\cite{ref23}                    & U                     & 14.467          & 0.634          & 0.434          & 6.810          & 16.430          & 0.658          & 0.275          & 5.879          & 13.738          & 0.265          & 0.428          & 5.050          \\
SCI~\cite{ref24}                     & U                     & 15.604          & 0.615          & 0.315          & 5.329          & 14.807          & 0.652          & 0.351          & 6.419          & 12.078          & 0.212          & 0.446          & \textbf{3.458} \\
PairLIE~\cite{ref37}                 & U                     & 18.137          & 0.720          & 0.324          & 5.338          & 18.499          & 0.835          & 0.248          & 4.031          & 17.241          & 0.360          & 0.328          & 4.137          \\ \hline
\textbf{TML (ours)}     & T                     & \textbf{19.358} & \textbf{0.743} & 0.360          & 4.727          & 20.780          & 0.820          & 0.303          & 5.311          & \textbf{19.444} & \textbf{0.460} & 0.351          & 5.318          \\ \hline
\end{tabular}
}
\caption{Quantitative comparisons with state-of-the-art methods on LOL and LSRW datasets. "U", "S", and "T" represent supervised learning, unsupervised learning and troublemaker learning respectively. $\uparrow$ represents that the larger the value, the more similar the prediction and reference are in color and brightness. $\downarrow$ represents that the smaller the value, the better the image quality.}
\label{tab01final}
\end{table*}

We first describe the implementation details, datasets, metrics and compared methods. Then, the quantitative and qualitative comparisons with existing methods are presented. We then conduct ablation experiments to prove the effectiveness of the design. We also perform visualization studies to demonstrate the explicability of TML and GDC. Finally, the limitations of our work are discussed. All experiments are completed on the NVIDIA 2080Ti GPU under PyTorch~\cite{ref51} framework.

\subsection{Experimental settings}
\label{sec:exp01}
{\bf Implementation details.} In the training phase, we use the AdamW~\cite{ref55} optimizer and the learning rate is set to 4e-5. The loss function is $smooth_{\mathcal{L}_1}$. TM and PM are trained for 15 epochs, and EM is trained for 30 epochs. The input size is $400\times{640}$ and the batch size is set to 8. We do not apply complex training techniques to verify the effectiveness of TML and GDC. The training process of both PM and EM exclusively utilizes normal-light images.

{\bf Datasets and metrics.} TM and EM are trained on LSRW~\cite{ref12}. In step 2, PM is trained on the BIGGER dataset we collected. The testing process is carried out on LSRW and LOL~\cite{ref11}. LSRW contains 5,600 low/normal-light image pairs for training and 50 pairs for testing. LOL has 485 and 15 pairs of data for training and testing, respectively. We select a total of 11,663 normal-light images from LSRW, SICE~\cite{ref52}, DPED~\cite{ref53}, and RELLISUR~\cite{ref54} to construct the BIGGER dataset. We employ peak signal-to-noise ratio (PSNR), structural similarity (SSIM)~\cite{ref58}, learned perceptual image patch similarity (LPIPS)~\cite{ref62}, natural image quality evaluator (NIQE)~\cite{ref63} to measure the effectiveness of the model. A higher PSNR/SSIM indicates the prediction is closer to the reference in color and brightness. The lower the LPIPS/NIQE, the higher the predicted image quality.

{\bf Compared methods.} We compare 13 recent state-of-the-art methods, including 8 supervised and 5 unsupervised methods. Supervised methods include Retinex-Net~\cite{ref11}, MBLLEN~\cite{ref14}, GLADNet~\cite{ref59}, KinD~\cite{ref19}, URetinex-Net~\cite{ref20}, SNR~\cite{ref65}, LLFlow~\cite{ref67}, Retinexformer~\cite{ref66} and unsupervised methods include EnlightenGAN~\cite{ref35}, RRDNet~\cite{ref36}, RUAS~\cite{ref23}, SCI~\cite{ref24}, PairLIE~\cite{ref37}. We use the officially provided codes and pretrained parameters for testing. To ensure fairness, the image sizes are all $400\times{640}$. All comparisons are completed on an NVIDIA 2080Ti GPU.

\subsection{Quantitative comparisons}
\label{sec:exp02}

\cref{tab01final} lists the quantitative evaluation results on the LOL, LSRW and RELLISUR~\cite{ref54} datasets. RELLISUR contains real low-light images for super-resolution. We chose 85 paired data from the validation set for testing. We can observe that the overall numerical performance of unsupervised methods is inferior to that of supervised methods. In the absence of references, unsupervised methods depend on priors such as complex loss functions to ensure model convergence, leading to limited generalization. In comparison, TML achieves competitive results. TML's good performance is attributed to the following two aspects. First, TML applies only normal-light images in step 2, which alleviates the dependence on paired data. Therefore, the training data can be easily expanded to improve the effect. Second, GDC can capture the correlations between elements over a larger area. UGDC can better capture semantic features in darker areas. TML excels in PSNR and SSIM on the LSRW and RELISSUR datasets, while Retinexformer performs better on LOL. We also observe that although TML performs competitively on PSNR and SSIM, it is suboptimal on LPIPS and NIQE. This proves that the prediction results perform well in terms of brightness and color, but there is room for improvement in image quality.

\subsection{Qualitative comparisons}
\label{sec:exp03}

\cref{fig06} shows qualitative comparisons between TML and other state-of-the-art methods. In terms of brightness, supervised methods are brighter overall than unsupervised methods. This is due to the high brightness of references in training data such as LOL and LSRW, so supervised methods can easily learn this brightness mapping. However, unsupervised methods lack reference guidance and rely on handcrafted priors to learn similar mappings, which is more difficult and lacks adaptability. TML is competitive in brightness because the input to step 2 is normal-light images, which means higher brightness is provided as a reference during the training stage. In terms of image quality, we observe that methods with good brightness and color performance do not necessarily have high image quality. Some methods (MBLLEN~\cite{ref14}, RRDNet~\cite{ref36}, SCI~\cite{ref24}) have mediocre brightness performance but higher image quality. TML makes trade-offs in brightness, color, and image quality. By slightly reducing image quality, it greatly improves brightness and color performance, as detailed in the ablation experiments.

\begin{table}[t]
\resizebox{1\columnwidth}{!}{%
\begin{tabular}{c|cccc|cccc}
\hline
\multirow{2}{*}{Method} & \multicolumn{4}{c|}{LOL}       & \multicolumn{4}{c}{LSRW}       \\ \cline{2-9} 
                        & PNSR$\uparrow$   & SSIM$\uparrow$  & LPIPS$\downarrow$ & NIQE$\downarrow$  & PNSR$\uparrow$   & SSIM$\uparrow$  & LPIPS$\downarrow$ & NIQE$\downarrow$  \\ \hline
200                     & 25.645 & 0.701 & 0.226 & \textbf{6.155} & \textbf{27.962} & \textbf{0.730} & \textbf{0.190} & 7.384 \\
400                     & 25.355 & 0.718 & 0.226 & 6.396 & 25.109 & 0.723 & 0.220 & 7.053 \\
1600                    & \textbf{27.512} & \textbf{0.753} & 0.219 & 6.477 & 25.269 & 0.727 & 0.242 & \textbf{6.915} \\
3200                    & 25.001 & 0.686 & 0.229 & 6.537 & 24.083 & 0.676 & 0.256 & 7.469 \\
5650                    & 27.376 & 0.742 & \textbf{0.205} & 6.591 & 25.588 & \textbf{0.730} & 0.222 & 7.586 \\ \hline
\end{tabular}%
}
\caption{TM's ablation study on the LSRW dataset. TM trained with only 200 paired data produces outputs close to real low-light images.}
\label{tab_TM_ablation}
\end{table}

\begin{table}[t]
\renewcommand\arraystretch{1.1}
\centering
\resizebox{\columnwidth}{!}{%
\begin{tabular}{c|c|c|cc|cccc}
\hline
\multirow{2}{*}{ID} & TM      & PM      & \multicolumn{2}{c|}{EM} & \multicolumn{4}{c}{LOL}                                            \\ \cline{6-9} 
                    & w/ GDC  & w/ GDC  & w/ GDC     & w/ res     & PSNR$\uparrow$            & SSIM$\uparrow$           & LPIPS$\downarrow$          & NIQE$\downarrow$           \\ \hline
A                   &         &         &            &            & 16.823          & 0.732          & 0.405          & 5.434          \\
B                   &         &         &            & $\surd$    & 19.072          & 0.814          & 0.323          & 5.315          \\
C                   & $\surd$ &         &            & $\surd$    & 19.201          & 0.812          & 0.332          & 6.202          \\
D                   &         & $\surd$ &            & $\surd$    & 19.301          & 0.806          & 0.335          & 5.716          \\
E                   &         &         & $\surd$    & $\surd$    & 20.622          & 0.820          & \textbf{0.294} & \textbf{4.792} \\
F                   & $\surd$ & $\surd$ & $\surd$    &            & 20.733          & 0.818          & 0.410          & 6.041          \\
G                   & $\surd$ & $\surd$ & $\surd$    & $\surd$    & \textbf{20.780} & \textbf{0.820} & 0.303          & 5.311          \\ \hline
\end{tabular}%
}
\caption{Quantitative results of ablation studies on the LOL dataset. The best results are shown in bold. $\surd$ in w/ GDC refers to adding GDC to U-Net, that is, using the UGDC model. Otherwise, U-Net is used. ${\surd}$ in w/ res indicates that EM predicts the $residual$ between $H$ and $H’$ in \cref{eq02}, rather than mapping from $H’$ to $H$ in \cref{eq01}.}
\label{tab02v2}
\end{table}

\subsection{Ablation study}
\label{sec:exp04}

We trained the TM with varying scales of paired data from LSRW. Notably, we compared the gap between TM’s outputs and real low-light images. \cref{tab_TM_ablation} shows that training TM with just 200 pairs yields outputs close to real low-light images. More paired training data does not significantly improve performance and may even reduce the quality of TM's predicted pseudo low-light images. Therefore, we ultimately selected 200 paired images from LSRW to train TM.

To verify the effectiveness of TML design, we conduct extensive ablation experiments, as shown in \cref{tab02v2}. In setting A, TM and PM use U-Net without GDC, and EM is removed. In setting B, TM, PM and EM all adopt U-Net. C, D, and E add GDC in TM, PM, and EM, respectively. In addition, EM from settings B to E all predict the residuals from $H'$ to $H$ in \cref{eq02}. Setting F adds GDC to TM, PM and EM. EM predicts a direct mapping of $H'$ to $H$ in \cref{eq01}. Setting G is what we ultimately adopted.

\begin{figure}[t]
  \centering
  
  \begin{subfigure}{0.24\linewidth}
    \includegraphics[width=0.98\linewidth]{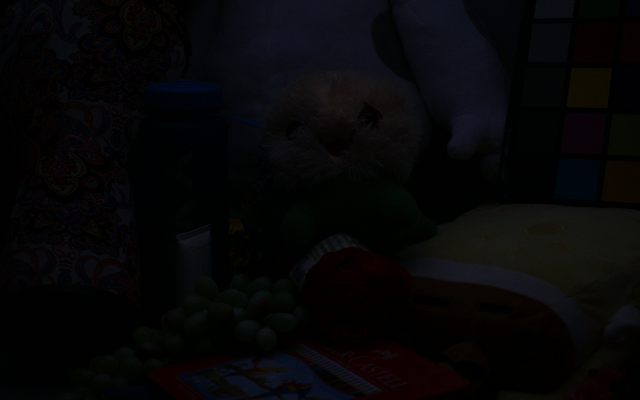}
    \caption*{Input}
    \label{fig08_1}
  \end{subfigure}
  \hfill
  \begin{subfigure}{0.24\linewidth}
    \includegraphics[width=0.98\linewidth]{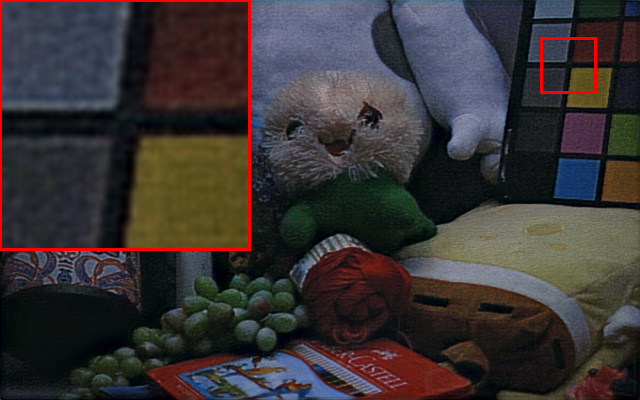}
    \caption*{A}
    \label{fig08_2}
  \end{subfigure}
  \hfill
  \begin{subfigure}{0.24\linewidth}
    \includegraphics[width=0.98\linewidth]{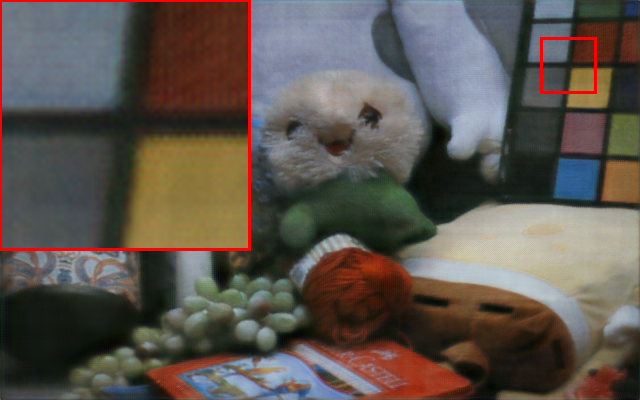}
    \caption*{B}
    \label{fig08_3}
  \end{subfigure}
  \hfill
  \begin{subfigure}{0.24\linewidth}
    \includegraphics[width=0.98\linewidth]{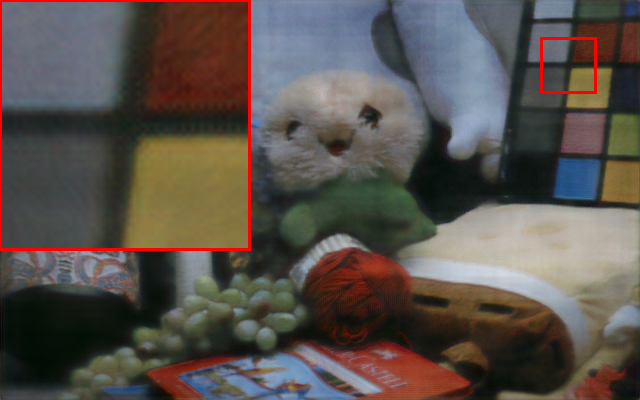}
    \caption*{C}
    \label{fig08_4}
  \end{subfigure}

  \begin{subfigure}{0.24\linewidth}
    \includegraphics[width=0.98\linewidth]{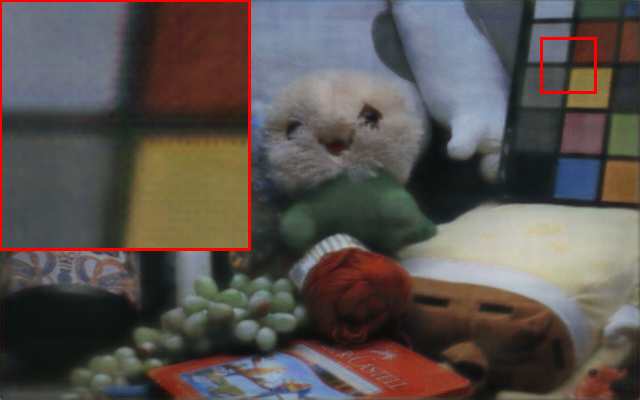}
    \caption*{D}
    \label{fig08_5}
  \end{subfigure}
  \hfill
  \begin{subfigure}{0.24\linewidth}
    \includegraphics[width=0.98\linewidth]{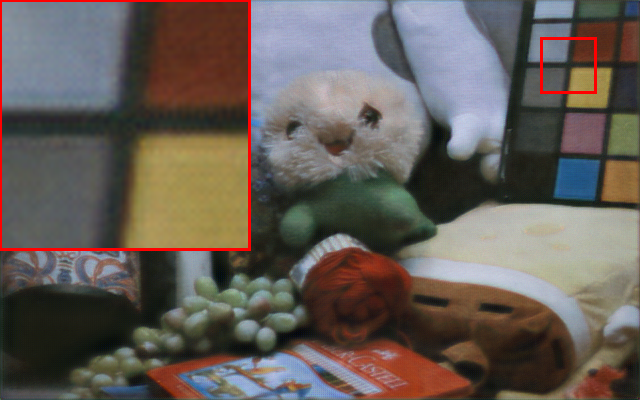}
    \caption*{E}
    \label{fig08_6}
  \end{subfigure}
  \hfill
  \begin{subfigure}{0.24\linewidth}
    \includegraphics[width=0.98\linewidth]{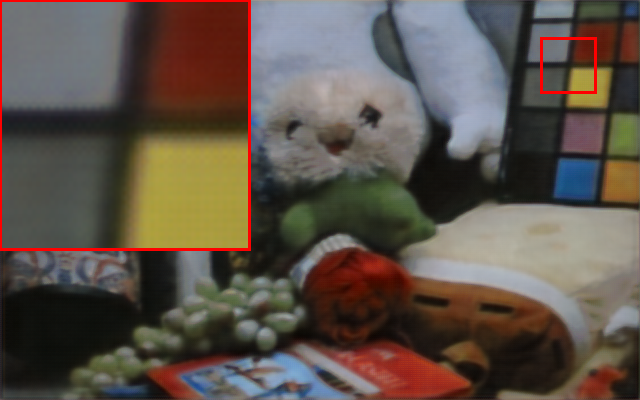}
    \caption*{F}
    \label{fig08_7}
  \end{subfigure}
  \hfill
  \begin{subfigure}{0.24\linewidth}
    \includegraphics[width=0.98\linewidth]{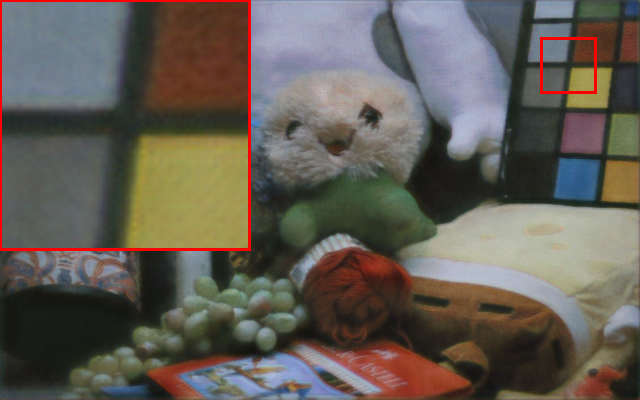}
    \caption*{G}
    \label{fig08_8}
  \end{subfigure}
  
  \caption{Visual comparisons of the ablation studies. Setting G achieves competitive performance in color, brightness and image quality.}
  \label{fig08}
\end{figure}

Comparing A and B, we observe that EM greatly improves the prediction performance in color, brightness (PSNR, SSIM) and image quality (LPIPS, NIQE). The comparison between B, C, D, and E proves that adding GDC to each model will improve the prediction performance in terms of color and brightness, proving the effectiveness of GDC and illustrating that capturing the correlation between elements is necessary for LLIE. Interestingly, adding GDC to TM and PM degrades LPIPS and NIQE, while adding it to EM improves LPIPS and NIQE, which indicates that GDC may slightly reduce sharpness. Setting G is slightly higher than F in PSNR and SSIM but has a larger improvement in LPIPS and NIQE. Although G is slightly lower than E in image quality, it performs better in color and brightness. According to a comprehensive comparison of settings A to G, we conclude that G makes a reasonable trade-off in color, brightness, and image quality. We visualize the predictions for settings A to G as illustrated in \cref{fig08}. The visual effect of A is poor. B, C, D, and F are average in terms of clarity. In comparison, the visual effect of G is slightly better than that of E, which is consistent with the previous analysis.

% Please add the following required packages to your document preamble:
% \usepackage{graphicx}
\begin{table}[t]
\centering
\resizebox{0.9\columnwidth}{!}{%
\begin{tabular}{c|c|cccc}
\hline
Dataset & Size  & PSNR$\uparrow$            & SSIM$\uparrow$           & LPIPS$\downarrow$          & NIQE$\downarrow$           \\ \hline
LOL     & 485   & 19.880          & 0.805          & 0.313          & 5.789          \\
LSRW    & 5600  & 20.546          & 0.817          & 0.312          & 5.634          \\
BIGGER  & 11663 & \textbf{20.780} & \textbf{0.820} & \textbf{0.303} & \textbf{5.311} \\ \hline
\end{tabular}%
}
\caption{Results for different sized datasets. Only normal-light images are used in the training phase. The results are obtained on the LOL test set. The best results are in bold.}
\label{tab03}
\end{table}

Because TML can easily increase training data, we exploit different sized datasets in step 2 during the training phase. The results are listed in \cref{tab03}. We observe that using larger datasets leads to better numerical results, proving that TML can improve prediction performance in color, brightness and image quality by increasing training data at a low cost. It is worth noting that although BIGGER does not contain LOL training data, it achieves the best performance on the LOL test set, indicating that TML has good generalization properties. All datasets exploit normal-light images.

\begin{table}[t]
\resizebox{1\columnwidth}{!}{%
\begin{tabular}{c|ccccc}
\hline
          & URetinex~\cite{ref20} & SNR~\cite{ref65}    & Retinexformer~\cite{ref66} & PairLIE~\cite{ref37} & TML (ours) \\ \hline
FLOPs(G)  & 227.631  & 90.425 & 66.478        & 87.294  & \textbf{58.670}     \\
Params(M) & 0.360    & 3.104  & 1.606         & \textbf{0.342}   & 4.442      \\ \hline
\end{tabular}%
}
\caption{Model FLOPs and parameter count comparison.}
\label{tab_FLOPs}
\end{table}

\subsection{FLOPs and parameters}
As shown in \cref{eq04_2,eq05_2,eq06_2,eq07_2}, GDC essentially conducts convolution operations with image patches as kernel parameters, thus its time complexity remains $O(n)$. We compared TML’s FLOPs and parameter count with several SOTA methods, shown in \cref{tab_FLOPs} (image size is $400 \times 640$). It is evident that TML has an advantage in FLOPs, indicating its computational efficiency. However, PairLIE~\cite{ref37} is more lightweight in terms of parameter count.

\subsection{Visualization study}
\label{sec:exp05}

\begin{figure}[t]

  \begin{subfigure}{0.32\linewidth}
    \includegraphics[width=0.98\linewidth]{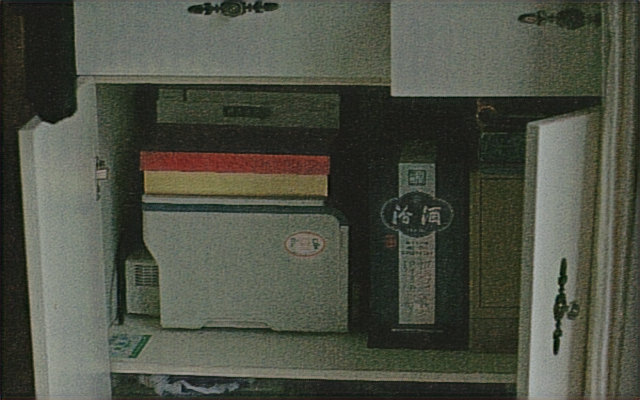}
  \end{subfigure}
  \hfill
  \begin{subfigure}{0.32\linewidth}
    \includegraphics[width=0.98\linewidth]{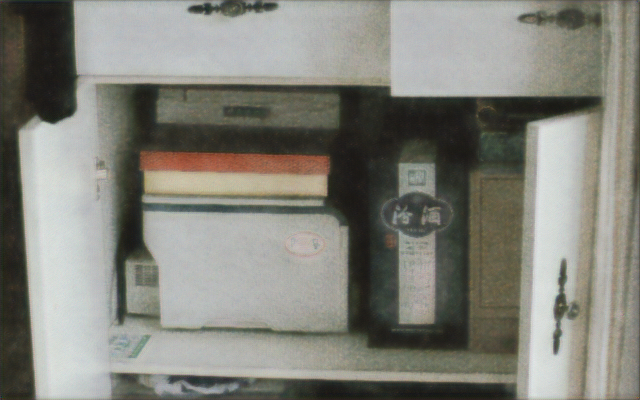}
  \end{subfigure}
  \hfill
  \begin{subfigure}{0.32\linewidth}
    \includegraphics[width=0.98\linewidth]{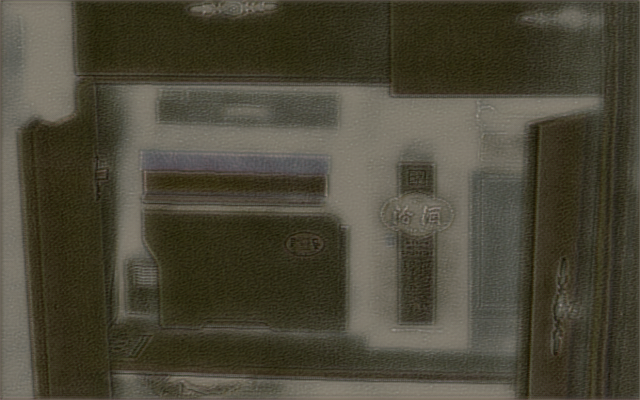}
  \end{subfigure}
  \vspace{0.5mm}

  \begin{subfigure}{0.32\linewidth}
    \includegraphics[width=0.98\linewidth]{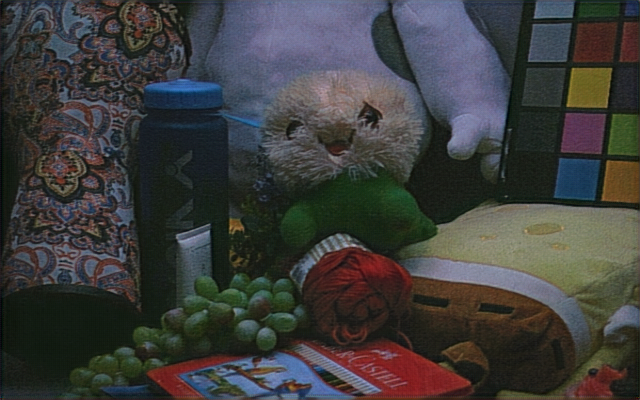}
  \end{subfigure}
  \hfill
  \begin{subfigure}{0.32\linewidth}
    \includegraphics[width=0.98\linewidth]{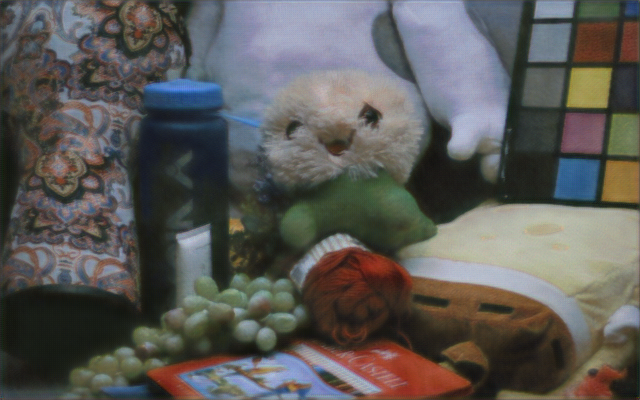}
  \end{subfigure}
  \hfill
  \begin{subfigure}{0.32\linewidth}
    \includegraphics[width=0.98\linewidth]{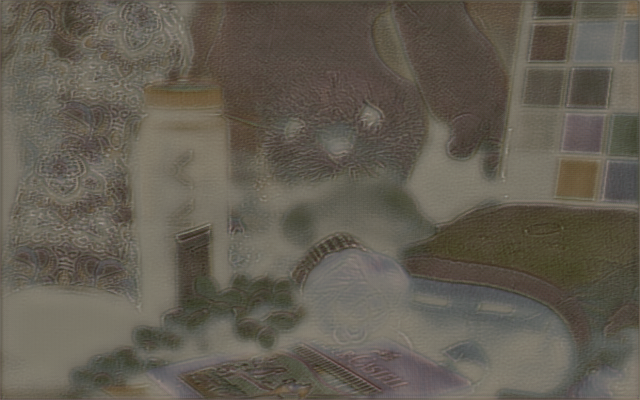}
  \end{subfigure}
  \vspace{0.5mm}

  \begin{subfigure}{0.32\linewidth}
    \includegraphics[width=0.98\linewidth]{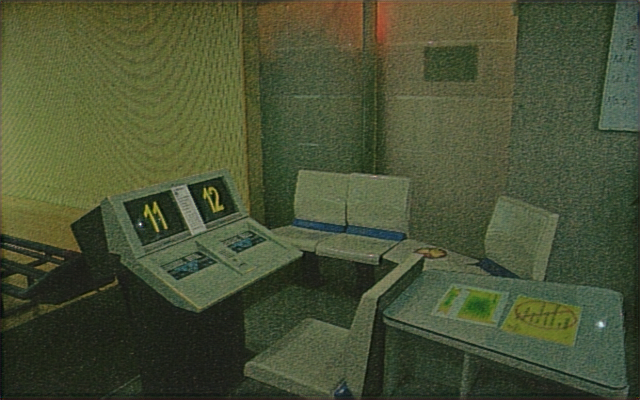}
    \caption*{Coarse $H'$}
  \end{subfigure}
  \hfill
  \begin{subfigure}{0.32\linewidth}
    \includegraphics[width=0.98\linewidth]{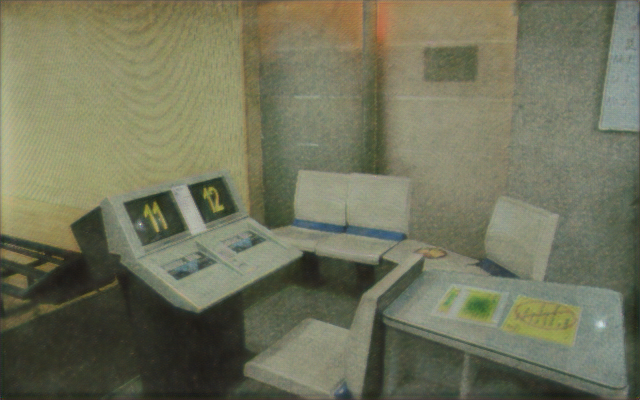}
    \caption*{Final $H$}
  \end{subfigure}
  \hfill
  \begin{subfigure}{0.32\linewidth}
    \includegraphics[width=0.98\linewidth]{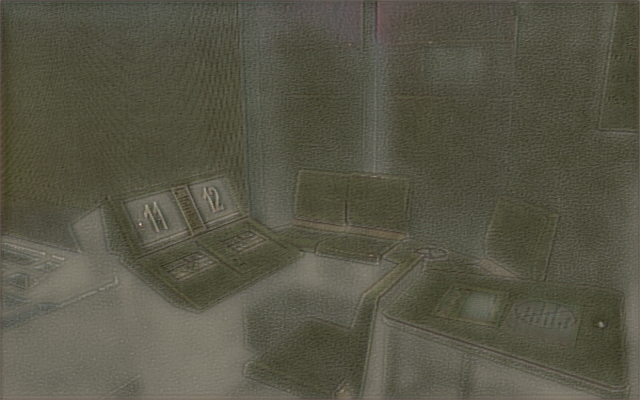}
    \caption*{Residual}
  \end{subfigure}
  \vspace{1mm}

  \caption{EM predictions. The first, second, and third columns represent $H'$, $H$, and $residual$, respectively, in \cref{eq02,eq03}.}
  \label{fig10}
\end{figure}

\textbf{Brightness residual.} We visualize the brightness residual predicted by EM as illustrated in \cref{fig10}. The first, second, and third columns represent $H'$, $H$, and $residual$, respectively, in \cref{eq02,eq03}. We observe that the $residual$ predicted by EM is an adaptive brightness map. Instead of brightening all elements of $H'$, it brightens the parts that should be brightened without changing the brightness of the remaining parts.

\begin{figure}[t]
  \centering

  \begin{subfigure}{0.24\linewidth}
    \includegraphics[width=0.98\linewidth]{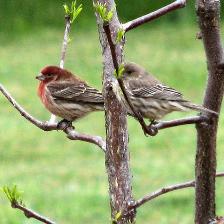}
  \end{subfigure}
  \hfill
  \begin{subfigure}{0.24\linewidth}
    \includegraphics[width=0.98\linewidth]{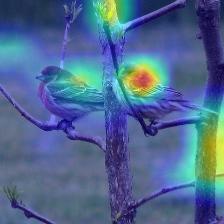}
  \end{subfigure}
  \hfill
  \begin{subfigure}{0.24\linewidth}
    \includegraphics[width=0.98\linewidth]{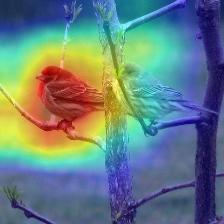}
  \end{subfigure}
  \hfill
  \begin{subfigure}{0.24\linewidth}
    \includegraphics[width=0.98\linewidth]{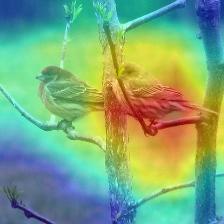}
  \end{subfigure}
  \vspace{0.5mm}

  \begin{subfigure}{0.24\linewidth}
    \includegraphics[width=0.98\linewidth]{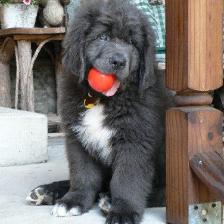}
    \caption*{Input}
  \end{subfigure}
  \hfill
  \begin{subfigure}{0.24\linewidth}
    \includegraphics[width=0.98\linewidth]{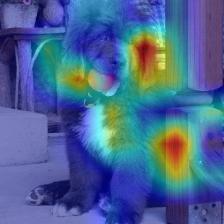}
    \caption*{ConvNeXt-Tiny}
  \end{subfigure}
  \hfill
  \begin{subfigure}{0.24\linewidth}
    \includegraphics[width=0.98\linewidth]{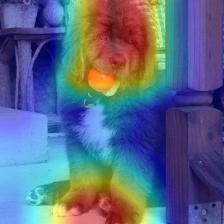}
    \caption*{ResNet18}
  \end{subfigure}
  \hfill
  \begin{subfigure}{0.24\linewidth}
    \includegraphics[width=0.98\linewidth]{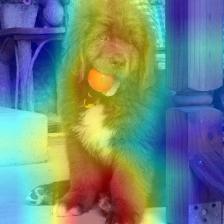}
    \caption*{ResGDC}
  \end{subfigure}
  \vspace{0.5mm}
  
  \caption{Grad-CAM for ConvNeXt-Tiny, ResNet18 and ResGDC. ResGDC can capture the correlation between a wider range of elements, indicating that GDC can improve the convolution module's ability to depict global element correlation.}
  \label{fig11}
\end{figure}

\textbf{GDC captures the correlations between elements on a larger scale.} We replace the convolutional layers of the last two layers in ResNet18~\cite{ref41} with GDC to form ResGDC. We then use the mini-ImageNet to train ConvNeXt-Tiny \cite{ref46}, ResNet18 and ResGDC with similar parameters and visualize their Grad-CAM~\cite{ref57}, as shown in \cref{fig11}. We observe that although ConvNeXt-Tiny performs better on ImageNet classification, it cannot characterize the correlations between elements well. Compared with ResNet18, ResGDC can capture a wider range of element correlations at the individual level. For instance, ResNet18 can capture information about only one bird, while ResGDC can capture information about two birds.

\subsection{Discussion}
\label{sec:exp06}

\begin{figure}[t]
  \centering

  \begin{subfigure}{0.24\linewidth}
    \includegraphics[width=0.98\linewidth]{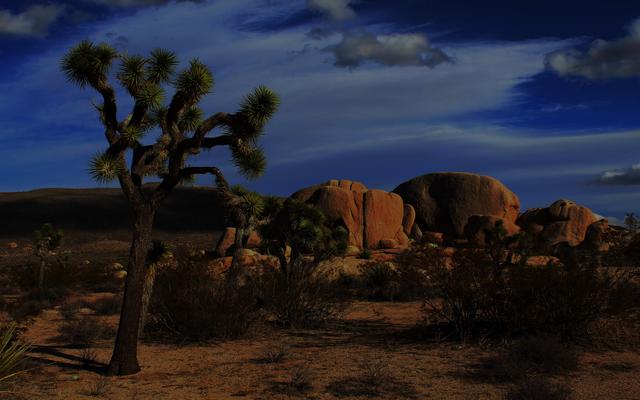}
    \caption*{Input}
  \end{subfigure}
  \hfill
  \begin{subfigure}{0.24\linewidth}
    \includegraphics[width=0.98\linewidth]{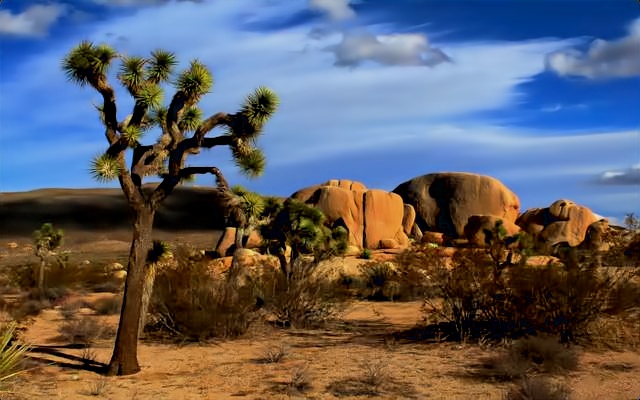}
    \caption*{MBLLEN}
  \end{subfigure}
  \hfill
  \begin{subfigure}{0.24\linewidth}
    \includegraphics[width=0.98\linewidth]{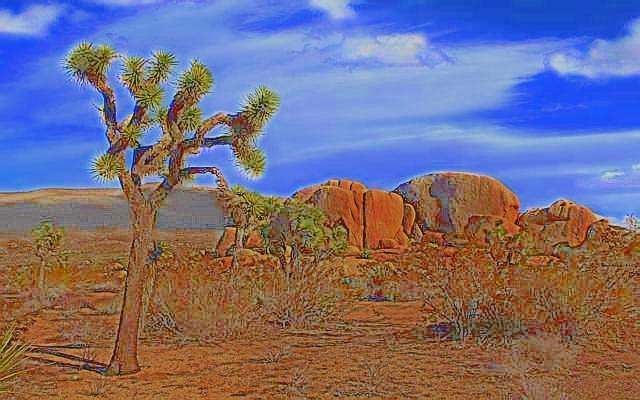}
    \caption*{Retinex}
  \end{subfigure}
  \hfill
  \begin{subfigure}{0.24\linewidth}
    \includegraphics[width=0.98\linewidth]{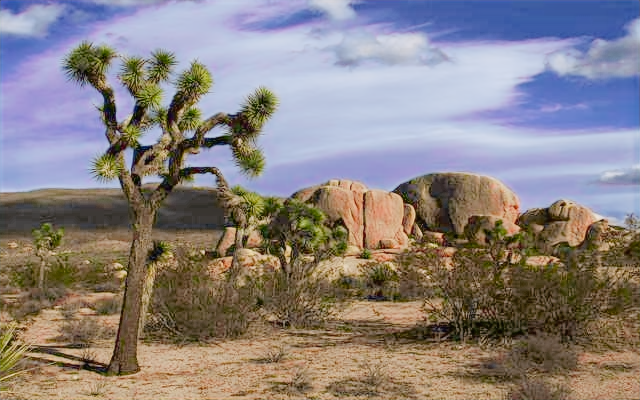}
    \caption*{GLAD}
  \end{subfigure}

  \begin{subfigure}{0.24\linewidth}
    \includegraphics[width=0.98\linewidth]{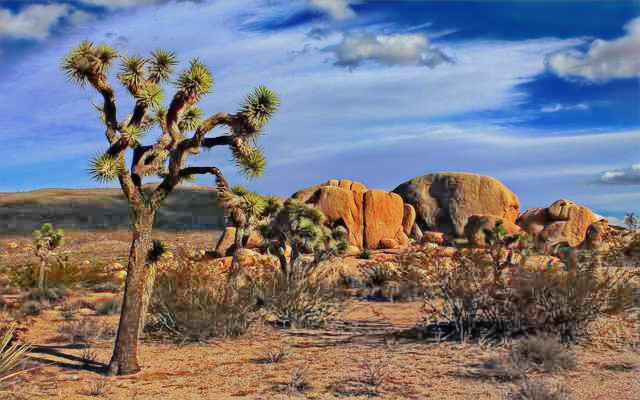}
    \caption*{KinD}
  \end{subfigure}
  \hfill
  \begin{subfigure}{0.24\linewidth}
    \includegraphics[width=0.98\linewidth]{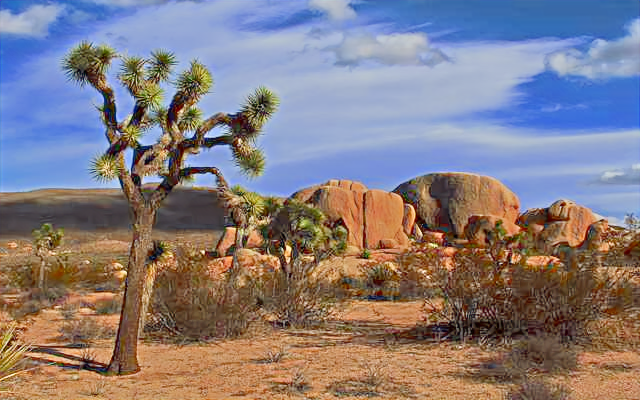}
    \caption*{URetinex}
  \end{subfigure}
  \hfill
  \begin{subfigure}{0.24\linewidth}
    \includegraphics[width=0.98\linewidth]{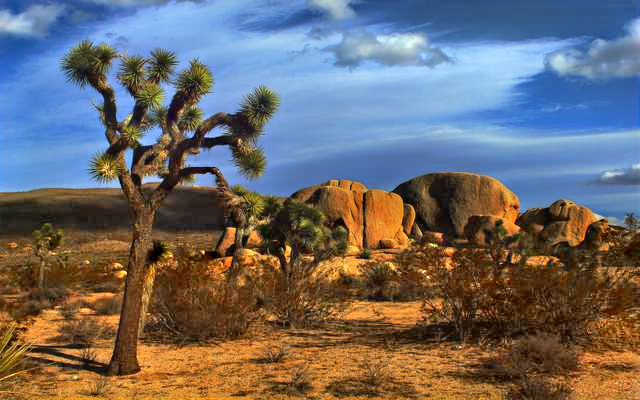}
    \caption*{EnGAN}
  \end{subfigure}
  \hfill
  \begin{subfigure}{0.24\linewidth}
    \includegraphics[width=0.98\linewidth]{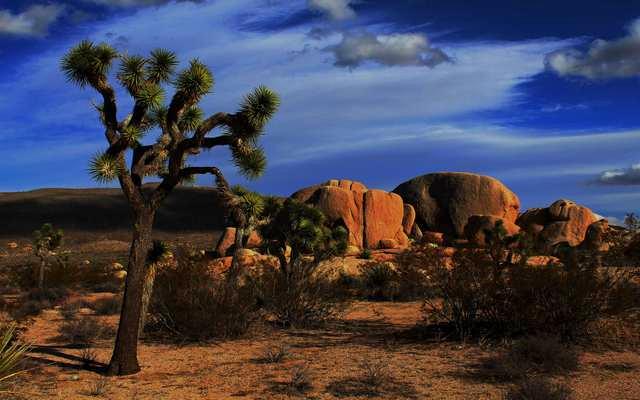}
    \caption*{RRDNet}
  \end{subfigure}

  \begin{subfigure}{0.24\linewidth}
    \includegraphics[width=0.98\linewidth]{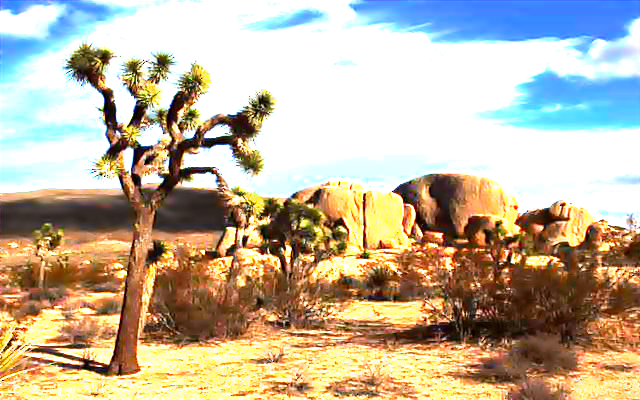}
    \caption*{RUAS}
  \end{subfigure}
  \hfill
  \begin{subfigure}{0.24\linewidth}
    \includegraphics[width=0.98\linewidth]{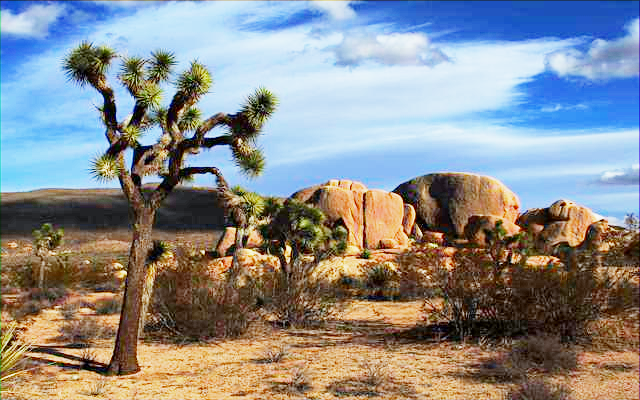}
    \caption*{SCI}
  \end{subfigure}
  \hfill
  \begin{subfigure}{0.24\linewidth}
    \includegraphics[width=0.98\linewidth]{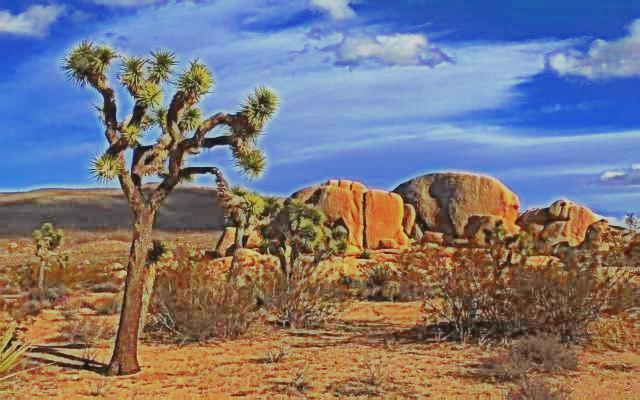}
    \caption*{PairLIE}
  \end{subfigure}
  \hfill
  \begin{subfigure}{0.24\linewidth}
    \includegraphics[width=0.98\linewidth]{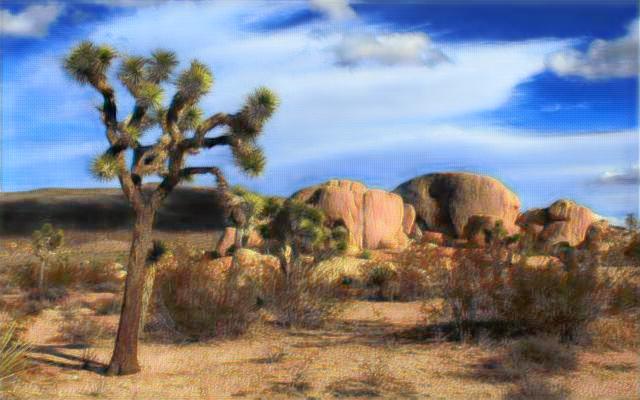}
    \caption*{TML (ours)}
  \end{subfigure}

  \caption{Performance of different methods on SICE dataset. TML has room for improvement in terms of sharpness and clarity.}
  \label{fig09v2}
\end{figure}

\textbf{Limitations.} TML has room for improvement in two aspects. First, despite alleviating the reliance on paired data, step 1 of TML still requires a small number of low/normal image pairs to train the pseudo low-light image generator. If the TM training process abandons paired data or uses an existing pretrained model as a pseudo low-light image generator, TML can be completely freed from the paired data limitations. Second, based on previous comparative experiments, we observe the TML image quality can still be optimized. \cref{fig09v2} shows the performance of different methods on the SICE~\cite{ref52} dataset. All methods have not been trained on this dataset. Although TML can obtain competitive visual effects, its sharpness and clarity are average.

\textbf{Future work.} The idea of TML is universal and can be applied to other scenarios that require paired data. In addition, GDC can capture correlations between a larger range of elements, so it can be applied in other general vision tasks. After solving the above two limitations, we will apply TML and GDC to other vision tasks.

\section{Conclusion}
\label{sec:con}

In this paper, we address two questions raised previously: First, we propose a training strategy called TroubleMaker Learning (TML). TML employs TM and PM separately to decrease and increase image luminance, and training data can be easily increased to improve the effect. It alleviates the dependence on pairwise data and uses a simple loss function. Second, we propose the Global Dynamic Convolution (GDC) module. GDC can capture element-wise correlations in a wide range with $O(n)$ time complexity. Extensive quantitative and qualitative experiments demonstrate the effectiveness of TML and GDC.

% --------------
{
    \small
    \bibliographystyle{ieeenat_fullname}
    \bibliography{main}
}

% WARNING: do not forget to delete the supplementary pages from your submission 
% \input{sec/X_suppl}

\end{document}